\documentclass[twocolumn]{aastex701}

\newcommand{\ha}{H$\alpha$}
\newcommand{\msun}{M_{\odot}}
\newcommand{\kms}{km\,s$^{-1}$}

\accepted{July 8, 2026}
\submitjournal{ApJL}

\shorttitle{The Shirui Group: A Compact Proto-group at $z \sim 5$}

\shortauthors{Laishram et al.}

\begin{document}

\title{A Compact Proto-group at $z \sim 5$: A Massive Galaxy Caught in Formation}

\correspondingauthor{Ronaldo Laishram}

\author[0000-0002-0322-6131]{Ronaldo Laishram}
\email{ronaldo.laishram@nao.ac.jp}
\affiliation{National Astronomical Observatory of Japan,
2-21-1 Osawa, Mitaka, Tokyo 181-8588, Japan}

\author[0000-0002-3801-434X]{Yusei Koyama}
\email{koyama.yusei@nao.ac.jp}
\affiliation{National Astronomical Observatory of Japan,
2-21-1 Osawa, Mitaka, Tokyo 181-8588, Japan}
\affiliation{Department of Astronomical Science,
The Graduate University for Advanced Studies,
2-21-1 Osawa, Mitaka, Tokyo 181-8588, Japan}

\author[0000-0002-4622-6617]{Fengwu Sun}
\email{fengwu.sun@cfa.harvard.edu}
\affiliation{Center for Astrophysics $|$ Harvard \& Smithsonian,
60 Garden St., Cambridge, MA 02138, USA}

\author[0000-0002-8512-1404]{Takahiro Morishita}
\email{morishita@astr.tohoku.ac.jp}
\affiliation{Astronomical Institute, Tohoku University,
6-3, Aramaki, Aoba, Sendai, Miyagi 980-8578, Japan}

\author[0000-0001-7440-8832]{Yoshinobu Fudamoto}
\email{yoshinobu.fudamoto@gmail.com}
\affiliation{Center for Frontier Science, Chiba University,
1-33 Yayoi-cho, Inage-ku, Chiba 263-8522, Japan}

\author[0000-0002-0479-3699]{Haruka Kusakabe}
\email{haruka.kusakabe.takeishi@gmail.com}
\affiliation{Department of General Systems Studies,
Graduate School of Arts and Sciences,
The University of Tokyo,
3-8-1 Komaba, Meguro-ku, Tokyo 153-8902, Japan}

\author[0000-0003-4337-6211]{Jakob M. Helton}
\email{jakobhelton@psu.edu}
\affiliation{Department of Astronomy \& Astrophysics,
The Pennsylvania State University, University Park, PA 16802, USA}

\author[0000-0001-6052-4234]{Xiaojing Lin}
\email{xiaojinglin.astro@gmail.com}
\affiliation{Department of Astronomy, Tsinghua University,
Beijing 100084, China}

\author[0000-0002-2993-1576]{Tadayuki Kodama}
\email{kodama@astr.tohoku.ac.jp}
\affiliation{Astronomical Institute, Tohoku University,
6-3, Aramaki, Aoba, Sendai, Miyagi 980-8578, Japan}

\author[0000-0003-1344-9475]{Eiichi Egami}
\email{egami@arizona.edu}
\affiliation{Steward Observatory, University of Arizona, 933 North Cherry Avenue, Tucson, AZ 85721, USA}

\author[0009-0009-8116-0316]{Kosuke Takahashi}
\email{kosuke.takahashi@astr.tohoku.ac.jp}
\affiliation{Astronomical Institute, Tohoku University,
6-3, Aramaki, Aoba, Sendai, Miyagi 980-8578, Japan}

\author[0009-0005-1487-7772]{Ryo Albert Sutanto}
\email{ryo.sutanto@astr.tohoku.ac.jp}
\affiliation{Astronomical Institute, Tohoku University,
6-3, Aramaki, Aoba-ku, Sendai, Miyagi 980-8578, Japan}

\author[0009-0009-3404-5673]{Novan Saputra Haryana}
\email{novan.haryana@astr.tohoku.ac.jp}
\affiliation{Astronomical Institute, Tohoku University,
6-3, Aramaki, Aoba-ku, Sendai, Miyagi 980-8578, Japan}

\author[0000-0003-1583-7404]{Zhengyi Chen}
\email{zhengyi.chen@nao.ac.jp}
\affiliation{National Astronomical Observatory of Japan,
2-21-1 Osawa, Mitaka, Tokyo 181-8588, Japan}

\begin{abstract}
We report the discovery of SCGG-z5, a compact galaxy proto-group at $z = 4.97$ in the MACS0416 field, identified from the SAPPHIRES Early Data Release. Six members are spectroscopically confirmed via \ha\ emission, spanning $4.96 \leq z_{\rm spec} \leq 4.98$ within a projected diameter of $\sim\!16$\,pkpc. Spectral energy distribution (SED) fitting yields individual stellar masses $8.4 \leq \log(M_*/\msun) \leq 9.8$, a total group stellar mass of $\log(M_*/\msun) = 10.07 \pm 0.04$; three of the six members lie above or on the star-forming main sequence at $z \sim 5$, by up to $0.5$\,dex. Pixel-by-pixel analysis reveals diverse resolved radial star-formation profiles: three members show declining specific SFR radial profiles and outward-rising stellar age gradients, consistent with inside-out stellar mass growth, while the most massive member shows a tentative inverted sSFR profile suggestive of reduced central star formation. The line-of-sight velocity dispersion over all six members is $\sigma_v = 375^{+55}_{-195}$\,\kms. The projected mass estimator yields $\log(M_{\rm PM}/\msun) \approx 12.30^{+0.30}_{-0.25}$, consistent with a dark-matter-dominated group halo. EAGLE simulations of structurally similar groups predict full coalescence by $z \sim 3$--$4$, with the merged remnant reaching $\log(M_*/\msun) > 11$ by $z \sim 1$, consistent with SCGG-z5 representing a rare pre-coalescence phase of early massive galaxy formation, possibly tracing the assembly of a future brightest group or cluster galaxy.
\end{abstract}


\section{Introduction}
\label{sec:intro}

Understanding the formation of the most massive galaxies (with a stellar mass of $M_* \gtrsim 10^{11}\,\msun$) is a central goal of modern observational cosmology. In the $\Lambda$CDM framework, these systems grow hierarchically: their dark matter haloes assemble through the successive merging of lower-mass subhaloes, with the most rapid stellar mass growth occurring at $z > 2$ through a combination of in situ star formation and the accretion of low-mass satellite galaxies \citep[e.g.,][]{Hopkins_et_al_2009, Oser_et_al_2010, Hirschmann_et_al_2012}. Galaxy overdensities, proto-clusters and proto-groups, trace the peaks of the primordial density field and are the preferred sites of this accelerated assembly. These structures are subject to downsizing, whereby galaxies in the densest environments, especially at the centre of the overdensity, evolve more rapidly than the field population \citep[e.g.,][]{Overzier_2016, Chiang_et_al_2017, Marrone_et_al_2018}. Within such environments, frequent gravitational interactions are expected to trigger starbursts, drive morphological transformations, and ultimately lead to group-scale coalescence into a single massive system \citep[e.g.,][]{MendesdeOliveira_Hickson_1994, Cluver_et_al_2013}.

Among high-redshift overdensities, compact galaxy groups (physically associated systems spanning only tens of kiloparsecs) represent a particularly short-lived and physically extreme phase. Dense, pre-virialized configurations enable efficient tidal torquing that strips angular momentum from the interstellar gas, driving massive inflows toward galaxy centres and triggering compaction: the rapid formation of ultra-dense, star-forming cores \citep{Dekel_Burkert_2014, Tacchella_et_al_2016}. Frequent mergers simultaneously remove gas from the outskirts and can fuel nuclear activity, accelerating gas consumption and potentially triggering quenching \citep[e.g.,][]{Hopkins_et_al_2011, Ellison_et_al_2011}. This fast-track evolutionary pathway makes compact groups plausible progenitors of the dense, quiescent systems observed at later cosmic epochs \citep[e.g.,][]{Ando_et_al_2022, Jin_et_al_2024, Haryana_et_al_2025}. Yet state-of-the-art cosmological simulations continue to underpredict the abundance of massive quiescent galaxies at $z > 4$ \citep[e.g.,][]{Girelli_et_al_2019, Carnall_et_al_2023, Valentino_et_al_2023, Baker_et_al_2025}, in part because direct observational constraints on the pre-coalescence merging phase remain elusive: imaging member galaxies before they merge requires both the sensitivity to detect low-mass, high-redshift systems and spectroscopic confirmation to establish genuine physical associations.

JWST has revealed a growing population of extreme galaxy overdensities and proto-cluster candidates spanning $z \sim 5$--$8$ \citep[e.g.,][]{Sun_et_al_2024, Helton_et_al_2024a, Helton_et_al_2024b, Morishita_et_al_2025a, Laishram_et_al_2026b}, demonstrating that dense large-scale environments were already shaping galaxy evolution within the first gigayear of cosmic history. Several compact multi-galaxy structures have been identified at high redshift \citep[e.g.,][]{Diaz-Santos_et_al_2018, Sillassen_et_al_2022, Jin_et_al_2023, Sillassen_et_al_2024, Arribas_et_al_2024, Morishita_et_al_2025b, Fudamoto_et_al_2025}, and JWST has now enabled their systematic study. CGG-z4 at $z = 4.3$ \citep{Brinch_et_al_2025} is spectroscopically confirmed with gas depletion times of $\lesssim 100$\,Myr, indicating imminent quenching. CGG-z5 at $z \sim 5.2$ \citep{Jin_et_al_2023} comprises six candidate members within $\sim\!10 \times 20$\,kpc$^2$, but membership rests on photometric redshifts alone. CGG-z7 at $z \sim 7.04$ \citep{Wei_et_al_2026} extends this family to the epoch of reionization as a pre-virialized structure near its first gravitational crossing. A related system at $z = 4.91$ \citep{Tanaka_et_al_2024}, with a clumpy morphology and central AGN, may represent a later stage in this evolutionary sequence as members begin to coalesce. Together, these systems suggest that compact, dynamically active multi-galaxy structures may be a recurring feature of massive galaxy assembly, yet the physical conditions governing their evolution at $z \sim 5$ are observationally largely unexplored.

In this Letter we report the discovery of SCGG-z5 (SAPPHIRES Compact Galaxy Group at $z \sim 5$, also referred to as the Shirui Group\footnote{Named after the Shirui Lily (\textit{Lilium mackliniae}), a rare flower found only in the Shirui Hills of Manipur, India.}), a compact galaxy group at $z = 4.97$ identified in the SAPPHIRES Early Data Release \citep{Sun_et_al_2025}. Six members are spectroscopically confirmed via \ha\ emission in JWST/NIRCam wide-field slitless spectroscopy, making this a rare compact group at $z \sim 5$ with all members confirmed, within a projected diameter of $\sim\!16$\,pkpc. The depth and 13-band photometric coverage of SAPPHIRES further enable pixel-by-pixel SED fitting of individual members, revealing a diversity of evolutionary stages within this single $\sim\!16$\,pkpc environment. The group dynamics are consistent with an early, dynamically assembling stage.

Throughout this work, we adopt a flat $\Lambda$CDM cosmology with $H_0 = 70$\,km\,s$^{-1}$\,Mpc$^{-1}$, $\Omega_{\rm m} = 0.3$, and $\Omega_\Lambda = 0.7$. All magnitudes are in the AB system.

\section{Data and Methods}
\label{sec:data}

\subsection{SAPPHIRES Early Data Release}
\label{subsec:sapphires}

This work uses data from the SAPPHIRES Early Data Release \citep{Sun_et_al_2025}, a JWST Cycle-3 Treasury programme (GO-6434; PI: Egami, E.) obtaining NIRCam imaging and wide-field slitless spectroscopy (WFSS) in pure parallel, with a total dual-channel exposure time of 47.2\,hours in the MACS~J0416 field.

NIRCam imaging covers 13 broad- and medium-band filters spanning $0.6$--$5.0$\,\micron\ (F070W, F090W, F115W, F140M, F150W, F182M, F200W, F210M, F277W, F335M, F356W, F410M, F444W), reaching $5\sigma$ point-source depths of $\sim$28.4--29.8\,AB mag \citep{Sun_et_al_2025}. WFSS observations in F356W and F444W Grism-C provide spectral coverage at $3.1$--$5.0$\,\micron; the median $5\sigma$ line sensitivity is $\sim$$6 \times 10^{-19}$\,erg\,s$^{-1}$\,cm$^{-2}$ at $3.7$\,\micron\ \citep{Sun_et_al_2025}. All images were reduced and drizzled to $0\farcs030$\,pix$^{-1}$ \citep{Sun_et_al_2025}.

Source detection and spectroscopic redshift measurement are described fully in \citet{Sun_et_al_2025}; here we summarise the aspects relevant to our sample. The photometric catalogue contains 22\,107 sources, of
which 1060 have confirmed spectroscopic redshifts at $z \simeq 0$--$8.5$. Redshift confidence levels (\texttt{zconf}) follow a 1--6 scale: \texttt{zconf}~$\geq 4$ indicates $\geq 2$ emission lines detected; \texttt{zconf}~$\leq 3$ indicates a single line.

\subsection{SED Fitting}
\label{subsec:sed}

We derive physical properties for all six members (Section~\ref{subsec:sample}) by fitting their
13-band NIRCam photometry with \textsc{Bagpipes} \citep{Carnall_et_al_2018},
adopting the same configuration as the SAPPHIRES EDR \citep{Sun_et_al_2025};
we refer the reader there for full details. Briefly, we use BPASS v2.2.1 binary stellar templates \citep{Stanway_Eldridge_2018} with a \citet{Kroupa_1993} IMF, a delayed-$\tau$ star-formation history (SFH), \citet{Calzetti_et_al_2000} dust attenuation with birth-cloud factor $\eta = 2$ for stars younger than 10\,Myr \citep{Sun_et_al_2025}, and free stellar metallicity and nebular ionisation parameter \citep{Hsiao_et_al_2023}. All parameters adopt uniform priors: stellar age $t \in [0.001, 2.0]$\,Gyr, $e$-folding timescale $\tau \in [0.01, 10.0]$\,Gyr, metallicity $Z \in [0.0005, 2.0]\,Z_\odot$, ionisation parameter $\log U \in [-4, -1]$, and dust attenuation $A_V \in [0, 8]$\,mag.
We adopt KRON apertures and apply point-source aperture corrections to determine total fluxes, with a 5\% photometric error floor \citep{Sun_et_al_2025}. Spectroscopic redshifts are held fixed (Table~\ref{tab:sample}). Posteriors are sampled with \texttt{Nautilus} \citep{Lange_2023}; all quantities are 50th-percentile values with 16th/84th uncertainties. Star-formation rates are averaged over the last 100\,Myr of the best-fit SFH.

For the spatially resolved analysis, NIRCam imaging in all 13 bands was PSF-matched to the F444W resolution ($\theta_{\rm FWHM} \approx 0\farcs106$) and adaptively binned per galaxy using \texttt{piXedfit} \citep{Abdurrouf_et_al_2021}, requiring $S/N \geq 3$ per bin in F444W. The same \textsc{Bagpipes} model configuration as the integrated fits was adopted, with spectroscopic redshifts held fixed. Given the extreme compactness of the group (minimum pair separation $0\farcs506$, corresponding to 16.9 pixels at the drizzled scale of $0\farcs030$\,pix$^{-1}$), bins are assigned strictly within each galaxy's segmentation footprint; those near adjacent members should be interpreted with caution.
Azimuthally averaged radial profiles were derived by computing the projected distance of each pixel from the SAPPHIRES EDR photometric catalogue position of each galaxy, converted to pixel coordinates via the WCS of the per-galaxy flux-map stamp; galaxy pixels are isolated using the \texttt{piXedfit} \texttt{GALAXY\_REGION} segmentation mask, ensuring each pixel contributes to at most one member's profile. Pixels are binned into eight equal-width circular annuli from zero to the maximum segmentation radius, with each bin reporting the median and 16th/84th-percentile values; bins containing fewer than five pixels are discarded. As a robustness check, profiles re-derived using the \texttt{PySersic} primary-component centroid positions in place of the catalogue coordinates yield consistent gradient classifications. To test whether the derived gradients are affected by low $S/N$ in the outskirts, this spatially resolved analysis was additionally repeated with $S/N \geq 5$ per bin; the outcome is presented alongside the radial profiles in Section~\ref{subsec:pixel}.

\section{Results}
\label{sec:results}

\subsection{Spectroscopic Confirmation and Physical Properties of SCGG-z5}
\label{subsec:sample}

\begin{figure*}[ht!]
\plotone{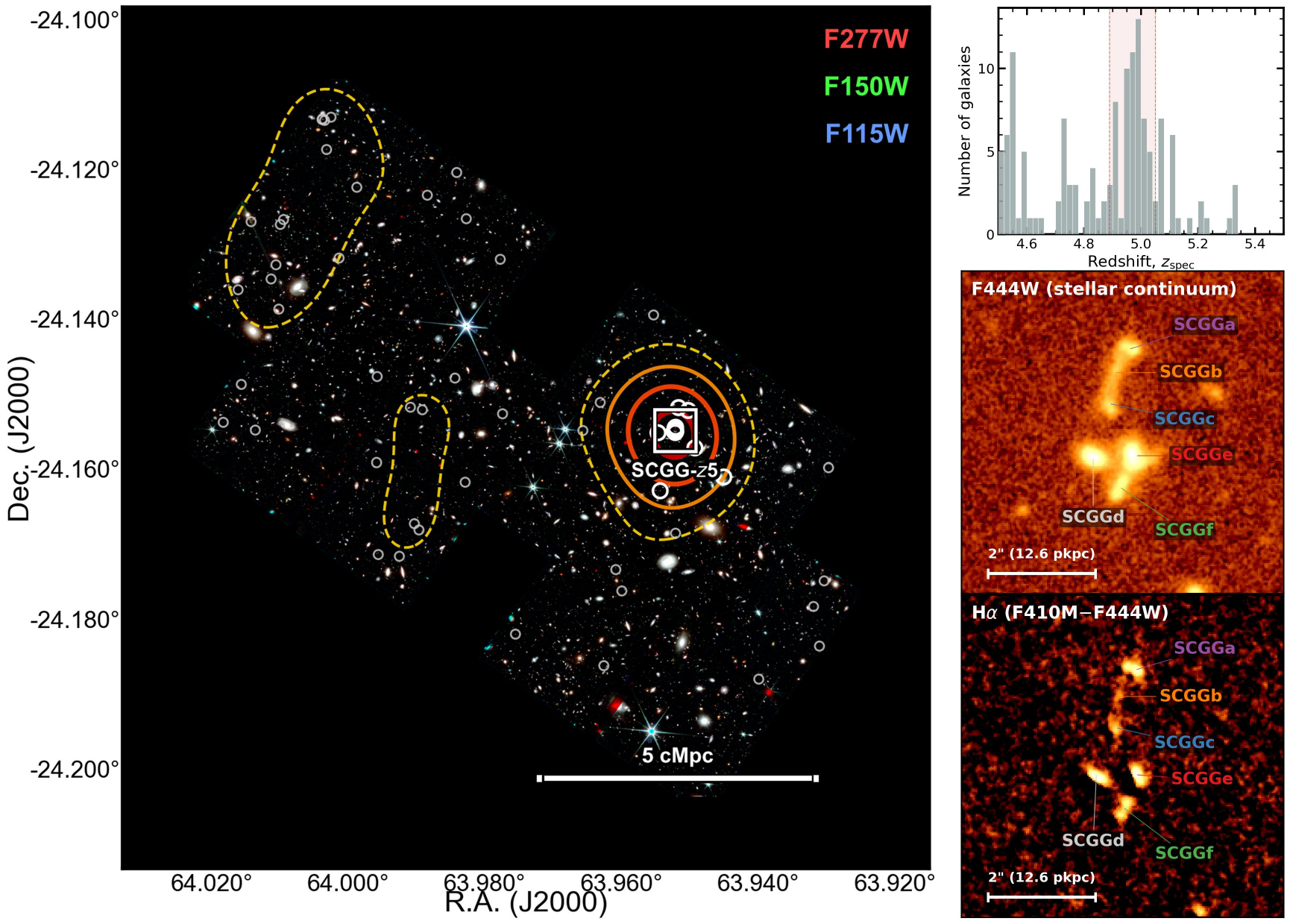}
\caption{
\textit{Left panel:} Wide-field JWST NIRCam RGB image (F277W/F150W/F115W). Overlaid contours show the 2D Gaussian KDE overdensity $\delta = (\rho/\langle\rho\rangle)-1$ of \ha\ emitters at $4.89 \leq z \leq 5.05$, with levels at $\delta = 1$ (gold dashed), 2 (orange), 3 (red), and 4 (dark red), computed with a fixed smoothing scale of $\sigma_{\rm KDE} = 1$\,cMpc. Open grey circles mark the 58 spectroscopic \ha\ emitters in the redshift slice; filled white circles indicate members within the $\delta \geq 2$ region. The white box marks the SCGG-z5 compact group.
\textit{Right panels:} \textit{Top:} Spectroscopic redshift distribution of SAPPHIRES \ha\ emitters at $4.5 \leq z_{\rm spec} \leq 5.5$; dashed lines and shading mark the overdensity spike window $4.89 \leq z \leq 5.05$ ($N_{\rm obs} = 58$, $7.1\sigma$). \textit{Middle:} F444W continuum (stellar continuum). \textit{Bottom:} \ha\ emission map from the F410M$-$F444W band difference. In all right panels, confirmed members are labelled by their SCGG designation (Table~\ref{tab:sample}).
\label{fig:rgb_ha}
}
\end{figure*}

We report SCGG-z5, a compact galaxy group at $z \approx 4.97$ in the MACS0416 field, identified through a systematic search for spectroscopic overdensities in the SAPPHIRES catalogue. A one-sided Poisson test applied to the 1D redshift distribution of \ha\ emitters at $4.89 \leq z \leq 5.05$ yields $N_{\rm obs} = 58$ versus $N_{\rm exp} = 19.1$ expected from the background rate measured over $3.9 \leq z \leq 6.6$ (excluding the spike window), corresponding to a significance of $7.1\sigma$ (one-sided Poisson $p$-value $p = 6.8 \times 10^{-13}$). The 1D redshift distribution illustrating this overdensity spike is shown in the top-right panel of Figure~\ref{fig:rgb_ha}.

To characterise the projected spatial distribution, we apply a 2D Gaussian kernel density estimate (KDE) to the positions of the 58 members, adopting a fixed smoothing scale of $\sigma_{\rm KDE} = 1$\,cMpc ($\approx 167$\,pkpc at $z \approx 4.97$), which is a factor of $\sim$21 larger than the maximum projected diameter of the compact group ($\approx 16$\,pkpc); this scale is chosen to trace large-scale structure and is not comparable to the group size. The KDE surface density field reaches $5.0\sigma$ above the field mean ($\langle\Sigma\rangle = 0.14$\,gal\,cMpc$^{-2}$; $\sigma$ denotes the standard deviation of the KDE density field) at its peak, with peak overdensity $\delta_{\rm peak} \equiv (\rho/\langle\rho\rangle) - 1 = 4.3$. As an independent, kernel-free verification, a Poisson test within a circular aperture of radius $R = 1$\,cMpc centred on the KDE peak yields $N_{\rm obs} = 11$ versus $N_{\rm exp} = 0.4$ ($7.0\sigma$, $p = 1.6 \times 10^{-12}$). Here $N_{\rm exp} = \Sigma_{\rm field} \times \pi R^2$, where $\Sigma_{\rm field}$ is the background H$\alpha$ surface density derived by dividing the expected background count from the 1D test above by the survey footprint area. This confirms that the overdensity significance is insensitive to the KDE smoothing scale. The KDE map is used solely to visualise the projected overdensity structure; all quoted significances are derived from the kernel-free Poisson tests above, which are independent of the KDE smoothing scale and not subject to KDE boundary effects. Tests with $R = 0.5$\,cMpc ($N_{\rm obs} = 8$, $N_{\rm exp} = 0.11$; $7.2\sigma$) and $R = 2.0$\,cMpc ($N_{\rm obs} = 15$, $N_{\rm exp} = 1.72$; $6.1\sigma$) demonstrate that the significance exceeds $6\sigma$ for all aperture radii tested.

The KDE density peak is located at R.A.\,=\,$63\fdg9527$, Decl.\,=\,$-24\fdg1556$, and the galaxy surface density within 1\,cMpc of this position is $\sim$26 times the field average. We selected all sources with spectroscopic redshifts in the range $4.89 \leq z_{\rm spec} \leq 5.05$ within a projected radius of $3\arcsec$ from this density peak; the unweighted geometric centroid of the six confirmed members, adopted as the group centre throughout, is R.A.\,=\,$63\fdg9522$, Decl.\,=\,$-24\fdg1547$. Six galaxies satisfy both criteria (Table~\ref{tab:sample}). Beyond the compact core, we identify one additional spectroscopically confirmed source within the group redshift range: ID\,14344 ($z = 4.975$, \texttt{zconf}~$= 6$, two emission lines) at a projected separation of $8\farcs3$ ($\approx\!52$\,pkpc) from the group centroid and $\Delta v \approx 255$\,\kms\ from the group mean redshift. Together with the eleven galaxies within 1\,cMpc of the KDE density peak that constitute the connected overdensity region, this suggests that the compact group is embedded within a broader large-scale structure at $z \approx 4.97$.

\begin{figure*}[ht!]
\plotone{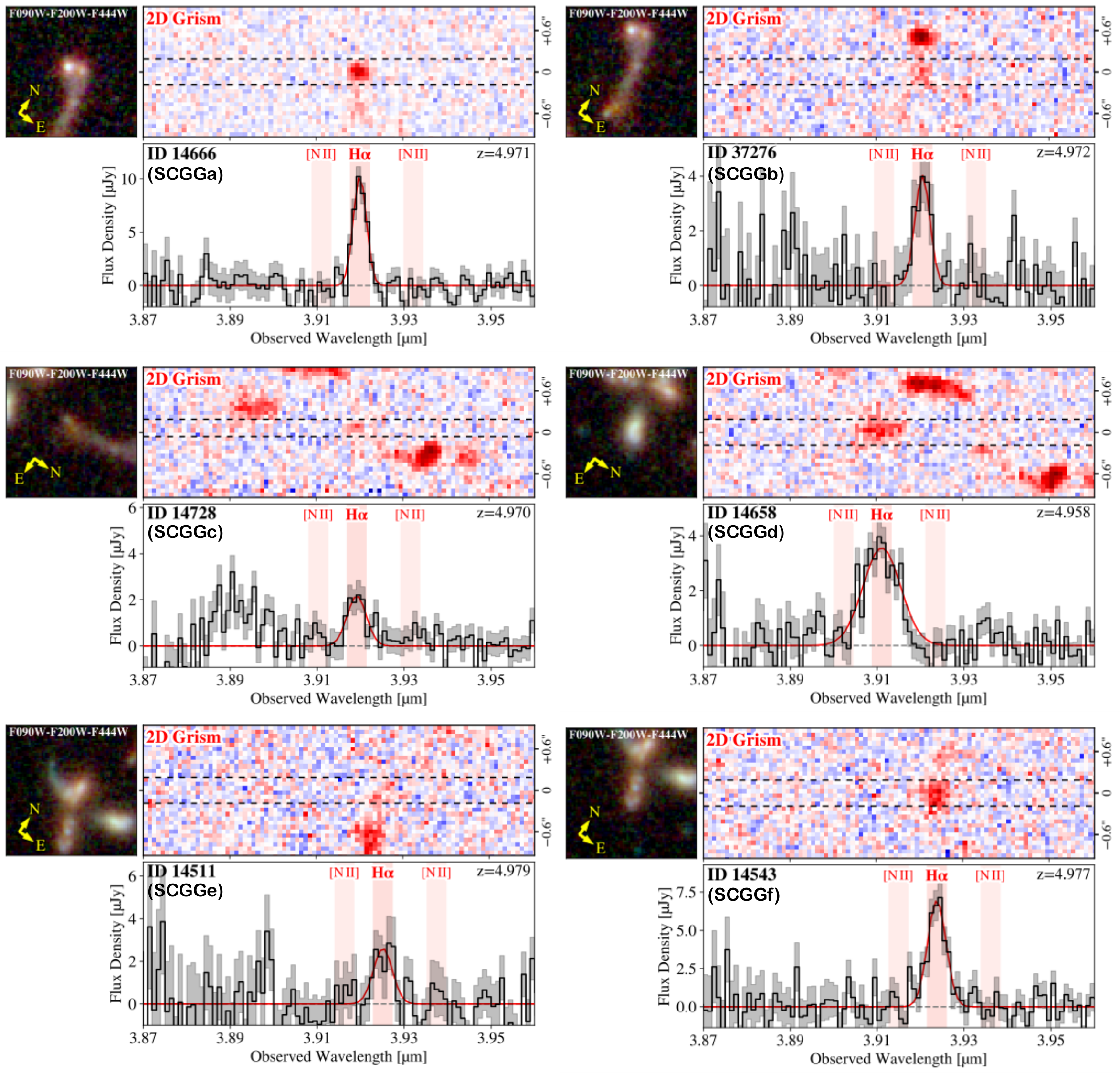}
\caption{
NIRCam grism spectra of the six SCGG-z5 members, ordered as in Table~\ref{tab:sample}. For each source: top-left shows the RGB image (F090W/F200W/F444W), and top-right shows the continuum-subtracted 2D grism spectrum. The bottom panel presents the extracted 1D spectrum (black), the best-fit \ha\ Gaussian (red), and the $1\sigma$ noise (grey). Pink bands indicate \ha\ (detections) and [N\,\textsc{ii}]$\lambda\lambda$6548,6583 (expected wavelengths). All six members show \ha\ with $3.7 \leq S/N \leq 15.0$.
\label{fig:grism_spectra}
}
\end{figure*}

Figure~\ref{fig:rgb_ha} presents the wide-field environment of SCGG-z5 overlaid with the KDE overdensity map (left panel), together with close-up imaging of the six members (right panels: spectroscopic redshift distribution, F444W continuum, and \ha\ emission map). The NIRCam grism spectrum of each member is shown in Figure~\ref{fig:grism_spectra}. The six members span spectroscopic redshifts $4.958 \leq z_{\rm spec} \leq 4.979$, corresponding to a line-of-sight velocity range of $\sim$1073\,\kms\ and a maximum projected separation of $2\farcs56$ ($\sim$16.1\,pkpc).

At $z \sim 5$, \ha\ falls at $\lambda_{\rm obs} \approx 3.93$\,\micron\ within the F356W and F444W grism bandpass, while H$\beta$ and [O\,\textsc{iii}] fall outside the covered wavelength range. SCGGb (\texttt{zconf}~$= 5$) shows two emission lines (\ha\ and [S\,\textsc{ii}]), satisfying the two-line criterion of the SAPPHIRES redshift confidence scale \citep[see][]{Sun_et_al_2025}. The [S\,\textsc{ii}]$\lambda$6731 line is detected at $\lambda_{\rm obs} = 4.020\,\micron$, consistent with $z = 4.972$. The remaining members are confirmed via single \ha\ detections (\texttt{zconf}~$= 1$--2), all with $S/N \geq 3$ across the available grism-filter combinations. Given the proximity in both sky position and line centroid, we interpret these six sources as members of a single compact galaxy group.

\begin{figure}[ht!]
\includegraphics[width=\columnwidth]{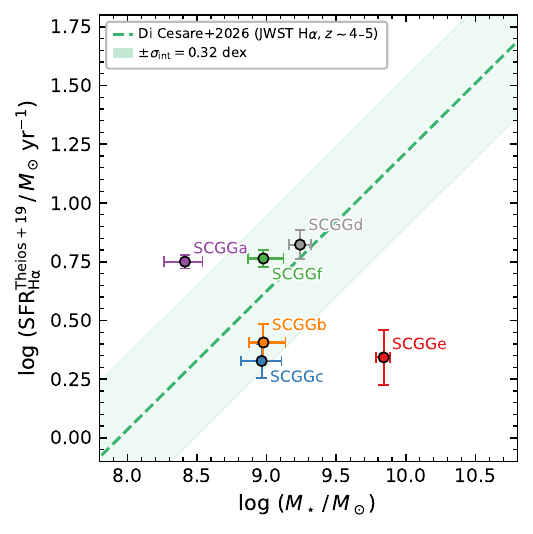}
\caption{
\ha-based SFRs (dust-corrected, \citealt{Theios_et_al_2019} calibration, Kroupa IMF) for the six proto-group members placed on the \citet{Di_Cesare_et_al_2026} H$\alpha$-based SFMS at $z\sim4$--5 (dashed green line; same calibration as our SFRs). The shaded green band shows the intrinsic scatter ($\pm\sigma_{\rm int} \approx 0.32$\,dex) of the \citet{Di_Cesare_et_al_2026} relation. Points are colour-coded by galaxy ID. SFRs are lower limits as no Balmer decrement is available
(Section~\ref{subsec:sample}).
\label{fig:sfr_ms}
}
\end{figure}

The SAPPHIRES F444W imaging ($5\sigma$ depth 29.1\,AB\,mag; \citealt{Sun_et_al_2025}) is complete to $\log(M_*/\msun) \approx 7.9$ at $z \sim 5$, estimated following \citet{Pozzetti_et_al_2010} as adopted by \citet{Paquereau_et_al_2025}. From \textsc{Bagpipes} SED fitting (Section~\ref{sec:data}), stellar masses span $8.42 \leq \log(M_*/\msun) \leq 9.84$, with a total group stellar mass $M_{*,{\rm tot}} = 10^{(10.07 \pm 0.04)}\,\msun$; all six members lie above the stellar mass completeness limit. \ha-based SFRs were computed using the \citet{Theios_et_al_2019} low-metallicity calibration ($\log C = 41.59$, adopted by \citealt{Di_Cesare_et_al_2026}) for a \citet{Kroupa_1993} IMF, with dust corrections from the SED-derived $A_V$ via the \citet{Calzetti_et_al_2000} attenuation law ($A_{\mathrm{H}\alpha} = 0.822\,A_V$). Since Balmer decrement measurements are unavailable at $z \sim 5$, the nebular excess attenuation \citep{Calzetti_2001} is not applied; the quoted SFRs are lower limits, as is standard for \ha-based estimates at this redshift. Placed on the \citet{Di_Cesare_et_al_2026} H$\alpha$-based SFMS at $z\sim4$--5 (Figure~\ref{fig:sfr_ms}; intrinsic scatter $\sigma_{\rm int} \approx 0.32$\,dex), three of the six members lie above or on the relation centre. SCGGa ($\Delta\mathrm{MS} = +0.47$\,dex) lies above the $1\sigma_{\rm int}$ band; SCGGf ($\Delta\mathrm{MS} = +0.15$\,dex) and SCGGd ($\Delta\mathrm{MS} = +0.06$\,dex) are consistent with the main sequence. SCGGe ($\Delta\mathrm{MS} = -0.78$\,dex) lies significantly below the main sequence (${\sim}2.4\,\sigma_{\rm int}$ below the relation). SCGGc ($\Delta\mathrm{MS} = -0.28$\,dex) and SCGGb ($\Delta\mathrm{MS} = -0.21$\,dex) lie below the relation centre but within the intrinsic scatter ($|\Delta\mathrm{MS}| < \sigma_{\rm int}$).

\begin{deluxetable*}{lcccccccc}
\tablewidth{0pt}
\tablecaption{Properties of the six spectroscopically confirmed members of SCGG-z5
\label{tab:sample}}
\tablehead{
  \colhead{ID} &
  \colhead{$z_{\rm spec}$} &
  \colhead{$\log(M_*/\msun)$} &
  \colhead{SFR$_{\mathrm{H}\alpha,\rm corr}$} &
  \colhead{SFR$_{\rm SED}$} &
  \colhead{$A_V$} &
  \colhead{Age} &
  \colhead{$r_{1/2}$} &
  \colhead{$n$} \\
  \colhead{} &
  \colhead{} &
  \colhead{} &
  \colhead{($\msun$\,yr$^{-1}$)} &
  \colhead{($\msun$\,yr$^{-1}$)} &
  \colhead{(mag)} &
  \colhead{(Gyr)} &
  \colhead{(kpc)} &
  \colhead{}
}
\startdata
14666 (SCGGa) & 4.971 & $8.42^{+0.13}_{-0.15}$ & $5.6^{+0.4}_{-0.4}$ & $3.0^{+1.1}_{-0.9}$ & $0.146^{+0.033}_{-0.031}$ & $0.065^{+0.035}_{-0.023}$ & $0.86^{+0.08}_{-0.07}$ & $0.72\pm0.05$ \\
37276 (SCGGb) & 4.972 & $8.98^{+0.16}_{-0.11}$ & $2.6^{+0.5}_{-0.4}$ & $2.8^{+1.3}_{-0.7}$ & $0.170^{+0.103}_{-0.085}$ & $0.89^{+0.17}_{-0.30}$ & $1.93^{+0.74}_{-0.54}$ & $0.77\pm0.10$ \\
14728 (SCGGc) & 4.970 & $8.96^{+0.13}_{-0.15}$ & $2.1^{+0.4}_{-0.3}$ & $5.0^{+2.5}_{-1.5}$ & $0.413^{+0.154}_{-0.124}$ & $0.38^{+0.32}_{-0.20}$ & $1.21^{b}\pm0.11$ & $0.81\pm0.13$ \\
14658 (SCGGd) & 4.958 & $9.24^{+0.08}_{-0.08}$ & $6.6^{+1.0}_{-0.9}$ & $20.2^{+4.5}_{-3.8}$ & $0.382^{+0.040}_{-0.045}$ & $0.052^{+0.030}_{-0.021}$ & $0.57\pm0.02$ & $0.70\pm0.04$ \\
14511 (SCGGe) & 4.979 & $9.84^{+0.04}_{-0.05}$ & $2.2^{+0.7}_{-0.5}$ & $22.2^{+3.5}_{-2.9}$ & $0.234^{+0.051}_{-0.040}$ & $0.76^{+0.19}_{-0.17}$ & $0.61\pm0.03$ & $1.85\pm0.18$ \\
14543 (SCGGf) & 4.977 & $8.98^{+0.14}_{-0.11}$ & $5.8^{+0.5}_{-0.5}$ & $8.6^{+1.1}_{-0.9}$ & $0.257^{+0.044}_{-0.046}$ & $0.18^{+0.11}_{-0.06}$ & $0.57^{a}\pm0.08$ & $1.2^{+0.4}_{-0.4}$ \\
\enddata
\tablecomments{
Source IDs from \citet{Sun_et_al_2025}; coordinates (R.A., Decl.) available therein.
Stellar masses, $A_V$, and ages: \textsc{Bagpipes} SED fitting (Section~\ref{sec:data}); 50th-percentile posteriors with 16th/84th uncertainties; age is the SFH onset time (delayed-$\tau$ model).
SFR$_{\mathrm{H}\alpha,\rm corr}$: \citet{Theios_et_al_2019} calibration with $A_{\mathrm{H}\alpha} = 0.822\,A_V$ \citep{Calzetti_et_al_2000}; lower limits (Section~\ref{subsec:sample}).
SFR$_{\rm SED}$: \textsc{Bagpipes} SFR from the delayed-$\tau$ SED fit; 16th/84th-percentile uncertainties.
$r_{1/2}$ and $n$: \texttt{PySersic} \citep{Pasha_Miller_2023} MCMC S\'ersic fitting in F200W (Section~\ref{subsec:morphology}); primary component unless otherwise noted; 16th/84th-percentile uncertainties.
$^{a}$~SCGGf: $r_{1/2}$ and $n$ refer to the secondary component ($1.44$\,kpc projected offset from primary).
$^{b}$~SCGGc: single-component parameters retained; two-component fit yields degenerate posteriors.}
\end{deluxetable*}

\subsection{Morphology}
\label{subsec:morphology}

We modelled all six members using \texttt{PySersic} \citep{Pasha_Miller_2023} with Markov Chain Monte Carlo (MCMC) sampling and empirical PSF (ePSF) models in the F200W band (rest-frame $\approx$3350\,\AA\ at $z \approx 4.97$). Four members (SCGGe, SCGGd, SCGGf, and SCGGa) required multiple S\'ersic components; SCGGe and SCGGf were each fitted with three components, and SCGGd and SCGGa with two components each. SCGGb was additionally fitted with two S\'ersic components to model its elongated, asymmetric F200W light distribution, with updated parameters listed in Table~\ref{tab:sample}. SCGGc was modelled with a single component; a two-component fit yields degenerate posteriors with overlapping effective radii and unconstrained S\'ersic indices (Table~\ref{tab:sample}, footnote~$b$). Results are shown in Figure~\ref{fig:pixel_sersic} (panel a).

All six members are resolved, with primary effective radii $r_{1/2} = 0.57$--$1.93$\,kpc ($1.6$--$5.4\times$ the F200W ePSF FWHM of $0\farcs057 \approx 0.36$\,kpc). The primary S\'ersic indices span $n = 0.68$--$1.85$, consistent with disk-like or irregular morphologies typical of star-forming galaxies at $z \sim 5$. The multi-component fits reveal physically motivated substructure: in SCGGe, the three-component fit reveals a prominent arc-like structure extending to the west of the main body and a fainter secondary component to the south-southwest, together consistent with tidal disturbance or stripped material; in SCGGf, photutils source detection independently confirms three distinct flux peaks within the galaxy, with the two secondary nuclei separated by $0.97$ and $1.19$\,kpc from the primary, consistent with a triple-nucleus system; in SCGGd, two components separated by 0.78\,kpc reveal a compact double-component morphology; in SCGGa, two compact nuclei are resolved; and in SCGGb, the two-component decomposition models its elongated, asymmetric stellar distribution. The best-fit S\'ersic parameters for all six members are listed in Table~\ref{tab:sample}.

Three of the six members (SCGGe, SCGGc, and SCGGd) show low-amplitude bipolar residual patterns within $1\,r_{1/2}$, suggestive of tidal perturbations. In all six members, the residual signal is stronger in the outer annulus ($r > r_{1/2}$) than within the half-light radius, plausibly associated with interactions in the compact group environment.

\subsection{Resolved Stellar Mass and Star Formation Maps}
\label{subsec:pixel}

Building on the morphological picture above, we performed pixel-by-pixel SED fitting to characterise the internal distribution of star formation within each member, following the procedure described in Section~\ref{subsec:sed}.

\begin{figure*}[ht!]
\includegraphics[width=\textwidth]{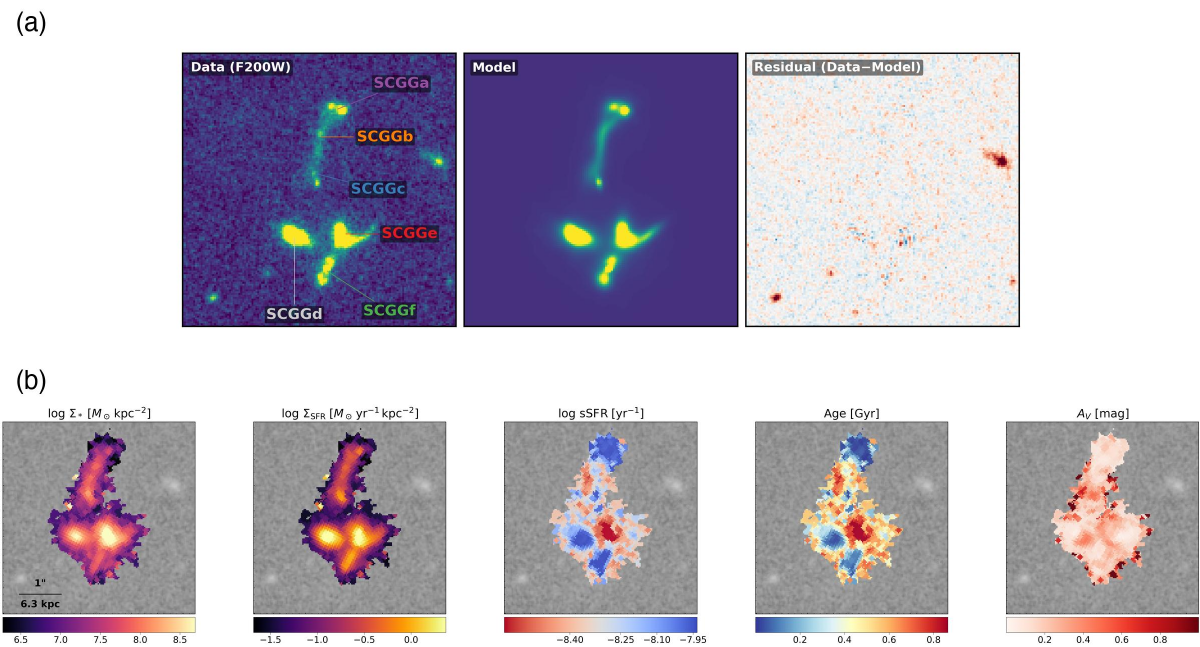}
\caption{
\textit{(a) Morphological decomposition} via multi-component S\'ersic fitting in F200W. From left to right: science image, best-fit model, and normalised residual. Residual structure and bipolar patterns in three of the six members are suggestive of tidal perturbations; the outer-dominated residuals common to all members are suggestive of group-scale tidal perturbations. \textit{(b) Resolved stellar properties} from pixel-by-pixel SED fitting. From left to right: stellar mass surface density ($\log \Sigma_{M_*}$, $M_\odot$\,kpc$^{-2}$), SFR surface density ($\log \Sigma_{\rm SFR}$, $M_\odot$\,yr$^{-1}$\,kpc$^{-2}$), specific SFR ($\log\,\mathrm{sSFR}$, yr$^{-1}$), mass-weighted age (Gyr), and dust attenuation ($A_V$, mag). Most members show centrally concentrated star formation with older stellar populations in the outskirts; SCGGe is the exception, with tentatively suppressed central sSFR suggestive of reduced central star formation \citep{Looser_et_al_2025}.
\label{fig:pixel_sersic}
}
\end{figure*}

\begin{figure}[ht!]
\includegraphics[width=\columnwidth]{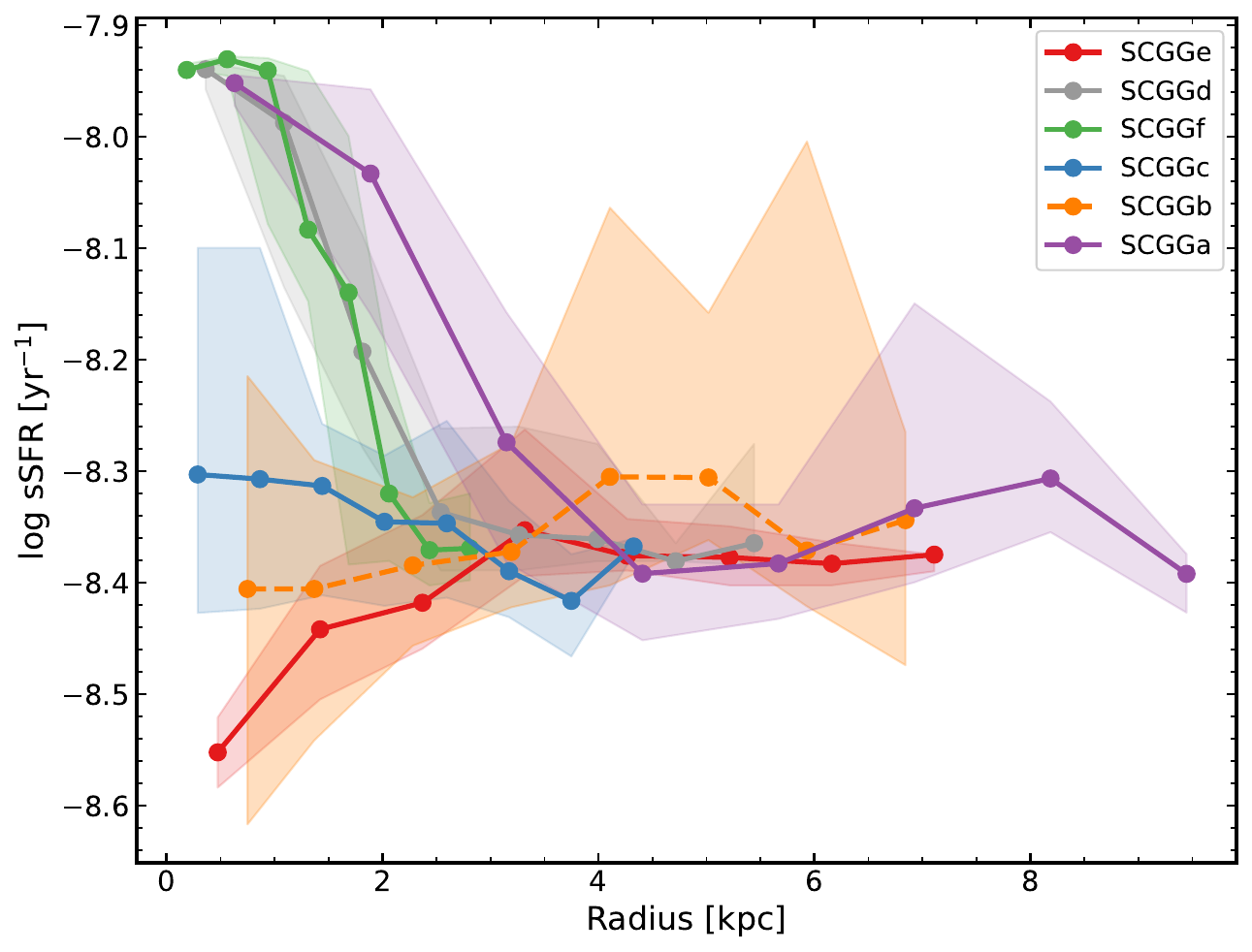}
\caption{
Azimuthally averaged $\log\,\mathrm{sSFR}$ radial profiles for the six proto-group members, derived from pixel-by-pixel \textsc{Bagpipes} SED fitting. Each galaxy is shown with a distinct colour (see legend); shaded bands indicate the $1\sigma$ bin-to-bin uncertainty. SCGGb (dashed line) has truncated central bins ($r < 0.85$\,kpc) due to its disturbed morphology. Three members (SCGGf, SCGGd, SCGGa) show clearly declining sSFR with radius (Spearman $\rho = -0.76$ to $-0.98$, $p < 0.05$); SCGGc shows a statistically significant but low-amplitude monotonic decline (Spearman $\rho = -0.929$, $p = 0.0009$; $\Delta\log\,\mathrm{sSFR} = 0.07$\,dex); SCGGb is flat. SCGGe (red) shows a tentative inverted profile, with lower sSFR in the centre than in the outskirts, suggestive of reduced central star formation \citep[see also][]{Looser_et_al_2025}.
\label{fig:radial_profiles}
}
\end{figure}

Resolved maps of $\log\Sigma_{M_*}$, $\log\Sigma_{\rm SFR}$, $\log\,\mathrm{sSFR}$, mass-weighted age, and $A_V$ are shown in Figure~\ref{fig:pixel_sersic} (panel b); azimuthally averaged radial profiles are presented in Figure~\ref{fig:radial_profiles}.
The maps reveal spatially distinct star-forming regions associated with individual members across the $\sim 16$\,pkpc extent of the group.
The most massive member, SCGGe, dominates the $\Sigma_{M_*}$ map yet shows lower central $\log\,\mathrm{sSFR}$ than the surrounding members and the oldest mass-weighted stellar age among the group, suggesting that differential star-formation histories are already operating within this compact system.

Three members (SCGGf, SCGGd, and SCGGa) show $\log\,\mathrm{sSFR}$ declining with radius, confirmed by the Spearman rank correlation test ($\rho = -0.76$ to $-0.98$; $p < 0.05$ in all cases), with the innermost radial bin sSFR exceeding the outermost by $\Delta\log\,\mathrm{sSFR} = 0.43$--$0.44$\,dex. Their mass-weighted stellar age profiles rise outward (younger centres, older outskirts; slopes $+0.05$ to $+0.21$\,dex\,kpc$^{-1}$), consistent with inside-out stellar mass growth.
SCGGc shows a statistically significant but low-amplitude declining sSFR profile (Spearman $\rho = -0.929$, $p = 0.0009$; $\Delta\log\,\mathrm{sSFR} = 0.07$\,dex), approximately six times smaller in amplitude than the three members above; its integrated SFR is consistent with the SFMS within the intrinsic scatter ($\Delta{\rm MS} = -0.28$\,dex, $|\Delta{\rm MS}| < \sigma_{\rm int}$), but the weak radial gradient amplitude does not constitute evidence for the pronounced centrally concentrated inside-out growth seen in SCGGf, SCGGd, and SCGGa. SCGGb shows a flat profile (Spearman $\rho = +0.68$, $p = 0.09$), consistent with its disturbed morphology (Section~\ref{subsec:morphology}).

SCGGe, the most massive member ($\log M_*/\msun = 9.84^{+0.04}_{-0.05}$), shows a tentative inverted sSFR profile (Spearman $\rho = +0.667$, $p = 0.07$; not statistically significant): the central bin ($r < 1$\,kpc) has $\log(\mathrm{sSFR}/\mathrm{yr}^{-1}) = -8.55^{+0.07}_{-0.04}$, lower than the outer mean of $-8.39^{+0.10}_{-0.06}$ ($\Delta\log\,\mathrm{sSFR} = -0.16$\,dex). It also hosts the highest central stellar mass surface density in the group ($\log\Sigma_{M_*} = 8.89^{+0.17}_{-0.42}\,\msun\,\mathrm{kpc}^{-2}$ within $r < 1$\,kpc), and its dust attenuation increases outward ($+0.056 \pm 0.015$\,mag\,kpc$^{-1}$, $p = 0.010$), suggesting the central region has lower dust content than the outer disk. Together, these radial signatures are suggestive of reduced central star formation in the most massive group member, consistent with a reduced cold-gas supply or a (mini-)quenching episode in the core \citep{Looser_et_al_2025}, while the outskirts remain active.

These radial trends are robust to the binning threshold: repeating the analysis with $S/N \geq 5$ per bin (Section~\ref{subsec:sed}) yields consistent gradient classifications. The declining sSFR profiles of SCGGf, SCGGd, and SCGGa retain statistical significance (Spearman $p < 0.05$ in all three cases) and SCGGb remains flat, confirming that the inside-out growth signatures are not driven by low-$S/N$ outskirt bins; SCGGe's tentative profile ($p = 0.07$ at $S/N \geq 3$) yields $p = 0.30$ at $S/N \geq 5$, consistent with its marginal significance at the fiducial threshold.

\subsection{Group Dynamics}
\label{subsec:dynamics}

The six spectroscopically confirmed members span a line-of-sight velocity range of $\Delta v_{\rm tot} = 1073$\,\kms\ relative to the group mean redshift $\bar{z} = 4.971$, from $\Delta v = -672$\,\kms\ (SCGGd)
to $\Delta v = +400$\,\kms\ (SCGGe). The velocity dispersion, computed as the unweighted standard deviation with $N-1$ degrees of freedom over all six members, is $\sigma_v = 375^{+55}_{-195}$\,\kms, where the asymmetric uncertainties are the 16th and 84th percentiles from $10^4$ bootstrap resamplings with replacement.
The wide confidence interval reflects the small sample size and is expected for $N = 6$; the large downward excursion is driven by bootstrapped subsamples that do not include the most kinematically extreme member.

The maximum projected pair separation is $2\farcs56$ (SCGGf to SCGGa), corresponding to $16.1$\,pkpc at $z \approx 4.97$. The minimum separation is $3.19$\,pkpc (SCGGb and SCGGa). The group half-projected radius, defined as the median projected separation of all six members from the unweighted geometric centroid, is
$R_{1/2} = 5.4$\,pkpc. The dynamical crossing time is $t_{\rm cross} \approx 14.1$\,Myr, only
$\sim$1.2\,per\,cent of the cosmic age at this epoch ($t_{\rm age}(z = 4.97) \approx 1.16$\,Gyr), indicating that the group members can complete multiple mutual passages within the current cosmic age.

We estimated the dynamical mass using the projected mass estimator of
\citet{Heisler_et_al_1985} (their Equation~11):
\begin{equation}
\label{eq:mpm}
M_{\rm PM} = \frac{f_{\rm PM}}{G\,(N - \alpha)}
             \sum_{i=1}^{N} v_i^2\,R_i ,
\end{equation}
where $v_i$ is the line-of-sight velocity of the $i$-th member relative to the group mean, $R_i$ is its projected separation from the unweighted geometric centroid, $G$ is the gravitational constant, $N = 6$, $f_{\rm PM}$ is a constant of proportionality that depends on the distribution of orbits \citep{Heisler_et_al_1985}, and $\alpha = 1.5$ is the correction for the offset between the arithmetic centroid of the tracers and the true centre of mass
\citep{Heisler_et_al_1985}. We adopted $f_{\rm PM} = 32/\pi$ for isotropic orbits, the value
\citet{Heisler_et_al_1985} recommend as giving the best agreement with numerical experiments. This yields $M_{\rm PM} = 2.0 \times 10^{12}$\,$\msun$ ($\log M_{\rm PM}/\msun = 12.30$; range $\approx 12.05$--$12.60$). Assuming purely radial orbits ($f_{\rm PM} = 64/\pi$) gives $M_{\rm PM} = 3.9 \times 10^{12}$\,$\msun$ ($\log M_{\rm PM}/\msun = 12.60$), which we treat as a conservative upper bound. To complement this with a lower bound, we use the N-body calibration of \citet{Heisler_et_al_1985}, who show that the projected mass estimator has an inter-quartile range of $\approx 0.5$\,dex for groups of $N \approx 5$--10 tracers; the statistical lower quartile is $\log M_{\rm PM}/\msun \approx 12.05$, giving a full confidence interval of $\log M_{\rm PM}/\msun \approx 12.05$--12.60.

The total stellar mass from \textsc{Bagpipes} SED fitting (Section~\ref{sec:data}) is $M_{*,{\rm tot}} = 1.2 \times 10^{10}$\,$\msun$ ($\log M_{*,{\rm tot}}/\msun = 10.07 \pm 0.04$), giving a stellar-to-dynamical mass fraction of
$f_* = M_{*,{\rm tot}}/M_{\rm PM} \approx 0.57$\,per\,cent ($\log f_* = -2.23^{+0.25}_{-0.25}$\,dex; the uncertainty is dominated by the \citealt{Heisler_et_al_1985} statistical scatter on $M_{\rm PM}$, $\pm 0.25$\,dex, while the propagated stellar mass uncertainty contributes only $\sim 0.04$\,dex in quadrature). This is consistent with the group being heavily dark-matter dominated; we discuss the comparison with the \citet{Behroozi_et_al_2019} stellar-to-halo-mass relation in Section~\ref{subsec:dynamics_discuss}.

\section{Discussion}
\label{sec:discussion}

\subsection{A Dynamically Young, Compact System}
\label{subsec:dynamics_discuss}

The kinematic and structural properties of SCGG-z5 are consistent with a compact, dynamically assembling system residing in a common dark-matter-dominated halo. With $t_{\rm cross} \approx 14.1$\,Myr (Section~\ref{subsec:dynamics}), the system is dynamically young, with members capable of completing multiple mutual passages within the current cosmic age. The stellar fraction $f_* \approx 0.57$\,per\,cent implies that the vast majority of the group mass resides in dark matter, with only a small fraction yet assembled into stars. We note that with $N = 6$ tracers both $\sigma_v$ and the projected mass
estimator carry substantial statistical uncertainty (bootstrap: $\sigma_v = 375^{+55}_{-195}$\,\kms; isotropic-to-radial range: $\log M_{\rm PM}/\msun = 12.05$--12.60); the physical conclusions are consistent across this range.

Beyond the statistical uncertainty, we note a systematic caveat. The projected mass estimator of \citet{Heisler_et_al_1985} was derived from the stellar hydrodynamic equation and calibrated on self-consistent equilibrium models; it therefore assumes the system is in a dynamical steady state. For a compact group at $z \approx 5$ that is still in an active assembly phase (as suggested by the evolutionary context discussed in this section), this assumption cannot be verified from photometric data alone. The values of $M_{\rm PM}$ and $\sigma_v$ should therefore be interpreted as characteristic of a dynamically assembling system rather than as equilibrium quantities; without spatially resolved kinematic data for all members, the direction and magnitude of any resulting bias cannot be determined.

The \citet{Behroozi_et_al_2019} stellar-to-halo-mass relation at $z \approx 5$ predicts $\log M_{\rm halo}/\msun = 12.00^{+0.08}_{-0.03}$ for a single galaxy with $\log M_*/\msun = 10.07\pm 0.04$. The \citet{Shuntov_et_al_2025} SHMR from COSMOS-Web abundance matching in the $4.5 < z < 5.5$ bin similarly predicts $\log M_{\rm halo}/\msun \approx 11.85$ for the same combined stellar mass, reflecting the elevated star-formation efficiency at $z > 3.5$ found in that study. Our directly measured $M_{\rm PM} = 2.0 \times 10^{12}$\,$\msun$ ($\log M_{\rm PM}/\msun = 12.30$; range $\approx 12.05$--$12.60$) exceeds both estimates by $\approx 0.3$--$0.45$\,dex. Both relations characterise the typical halo of a single field galaxy, whereas the dynamical mass of SCGG-z5 reflects the collective group halo enclosing six member galaxies; the excess is therefore consistent with SCGG-z5 occupying a genuine group-scale dark-matter halo rather than representing an anomalously overmassive single system.

As a cross-check under a virialization assumption, we apply the dark-matter velocity dispersion--mass scaling of \citet{Evrard_et_al_2008} (their Equation~6), $\sigma_{\rm DM} = 1082.9\,[h(z)\,M_{200}/10^{15}\,M_{\odot}]^{0.3361}\,{\rm km\,s^{-1}}$, with $h(z=4.97)=5.62$. Treating $\sigma_v$ as a proxy for $\sigma_{\rm DM}$ (velocity bias $b_v \approx 1.00\pm0.05$; \citealt{Evrard_et_al_2008}) and inverting for $\sigma_v = 375$\,\kms\ yields $M_{\rm vir} \approx 7.6\times10^{12}$\,$\msun$ ($\log M_{\rm vir}/\msun \approx 12.9$), a factor of $\approx\!4$ above $M_{\rm PM}$. Equivalently, evaluating this relation at $M_{\rm PM}$ predicts $\sigma_{\rm DM} \approx 240$\,\kms, a factor of $\approx\!1.6$ below the observed $\sigma_v$. The \citet{Evrard_et_al_2008} scaling is demonstrated to hold self-similarly over $10^{10}$--$10^{15}\,h^{-1}\,\msun$ (their Figure~6); the offset between $M_{\rm vir,Evrard}$ and $M_{\rm PM}$ is therefore consistent with $\sigma_v$ being elevated above the equilibrium expectation in a dynamically assembling system in which the kinetic energy has not yet settled to its virial value.

Comparable compact groups at similar and earlier epochs are discussed in Section~\ref{subsec:analogues}; here we focus on the dynamical implications of the SCGG-z5 measurements themselves. The relatively high velocity dispersion of SCGG-z5 is consistent with a dynamically active, multi-passage assembly stage rather than a first-crossing configuration.

By analogy with EAGLE hydrodynamical simulations of CGG-z5-like compact groups \citep{Jin_et_al_2023}, such structures are expected to coalesce into a single galaxy by $z \sim 3$--$4$, continuing to grow to $\log M_*/\msun > 11$ by $z \sim 1$. If SCGG-z5 follows a similar evolutionary path, these observations are consistent with catching a short-lived, pre-coalescence phase of massive galaxy formation \citep{Jin_et_al_2023}. The mean stellar age of the six members from \textsc{Bagpipes} SED fitting, $\langle t_{\rm gal}\rangle \approx 0.39$\,Gyr, indicates that star formation in this system began at $z \gtrsim 7$. Adding the $\sim\!400$\,Myr merger timescale from the EAGLE analogue structures \citep{Jin_et_al_2023} suggests an indicative assembly timescale from star-formation onset to coalescence of $\langle t_{\rm gal}\rangle + t_{\rm merger} \approx 0.8$\,Gyr, consistent with SCGG-z5 being observed in the early formation phase of a system that will appear as a massive galaxy by $z \sim 3$--4 \citep{Valentino_et_al_2020}. We note that stellar ages from a parametric delayed-$\tau$ SFH carry model-dependent systematic uncertainties arising from age, metallicity, and dust degeneracies that are not captured by the posterior errors; stellar masses are substantially more robust to the assumed SFH. The inference that star formation began at $z \gtrsim 7$ is derived from the \textsc{Bagpipes} best-fit SFH-onset age $\langle t_{\rm gal}\rangle \approx 0.39$\,Gyr, but the exact epoch and the assembly timescale above should be treated as indicative.

\subsection{Morphological Interactions and Resolved Star Formation}
\label{subsec:morph_context}

The morphological and resolved star-formation analysis presented in Sections~\ref{subsec:morphology} and \ref{subsec:pixel} reveals a system in which gravitational interactions and mass-dependent evolution are
already operating simultaneously at $z \approx 5$.

The morphological disturbances quantified in Section~\ref{subsec:morphology} are suggestive of tidal interactions within the group. Three of the six members show bipolar residual patterns within $1\,r_{1/2}$, and the outer-dominated residual signal is common to all six, suggesting that tidal perturbations are pervasive across the group. Observational evidence for tidal gas stripping on group scales comes from ALMA [CII] observations of a compact merging system at $z \sim 4.57$ \citep{Ginolfi_et_al_2020}, in which $\sim$50\,per\,cent of the total [CII] luminosity resides in a circumgalactic envelope extending to $\sim$30\,pkpc, attributed to gas stripped by strong gravitational interactions. At such small projected separations ($\lesssim 16$\,pkpc), the compact configuration may enable efficient tidal torquing: angular momentum is stripped from the interstellar gas, driving inflows toward galaxy centres and potentially triggering compaction (the formation of ultra-dense, star-forming cores; \citealt{Dekel_Burkert_2014, Tacchella_et_al_2016}).

Three members (SCGGf, SCGGd, and SCGGa) show declining sSFR radial profiles, with their central regions forming stars at higher specific rates than their outskirts (Section~\ref{subsec:pixel}).
This inside-out growth pattern is consistent with the compaction scenario, where tidal torques funnel cold gas toward galaxy centres, building concentrated stellar cores at early epochs \citep[e.g.,][]{Tacchella_et_al_2016, Tacchella_et_al_2018}. Observational support comes from integral field spectroscopy (IFS) studies of nearby interacting galaxies, which find a factor of $\sim\!2$--3 enhancement of central specific star formation relative to non-interacting controls, attributed to tidally induced gas inflows \citep{Barrera-Ballesteros_et_al_2015}. The simultaneous occurrence of this signature in three of the six group members suggests that the compact environment has, in these galaxies,
enhanced centrally concentrated star formation rather than disrupting it; each of these members appears to be in an active central mass-assembly phase plausibly linked to interaction-driven gas inflow.
SCGGc also shows a Spearman-significant monotonically declining profile ($\rho = -0.929$, $p = 0.0009$), but its amplitude is only $\Delta\log\,\mathrm{sSFR} = 0.07$\,dex (approximately six times smaller than the three members above), and its integrated SFR is consistent with the SFMS within the intrinsic scatter ($\Delta{\rm MS} = -0.28$\,dex, $|\Delta{\rm MS}| < \sigma_{\rm int}$); we therefore interpret its radial gradient as physically marginal and do not include it in the inside-out growth population.

The most massive member, SCGGe ($\log M_*/\msun = 9.84^{+0.04}_{-0.05}$), presents a qualitatively different picture: its tentative inverted sSFR radial profile (Section~\ref{subsec:pixel}) shows central star formation suppressed relative to the outer disk ($\log(\mathrm{sSFR}_{\rm cen}/\mathrm{yr}^{-1}) = -8.55^{+0.07}_{-0.04}$, within $\lesssim 0.5$\,dex of the SFMS). We caution that this does not unambiguously indicate long-term quenching; at $\log M_*/\msun \approx 9.84$ and $z \approx 5$, such suppression is consistent with a transient (mini-)quenching episode driven by a temporary disruption of cold-gas inflow \citep{Looser_et_al_2025}, or with inside-out stellar mass growth reducing the central gas fraction \citep{Tacchella_et_al_2016}. In the compaction scenario, the nuclear stellar concentration exhausts the central gas supply while the outer disk remains temporarily active; direct observational evidence for this mode has been found in compact star-forming galaxies at $z \sim 2$ \citep{Spilker_et_al_2019}. A similar qualitative pattern has also been seen in spatially resolved studies of massive protocluster galaxies at lower redshift, where the inferred radial structure is consistent with inside-out quenching, although SCGG-z5 probes a much earlier and more compact pre-coalescence stage \citep{Laishram_et_al_2026a}. The high central stellar mass surface density ($\log\Sigma_{M_*} = 8.89^{+0.17}_{-0.42}\,\msun\,\mathrm{kpc}^{-2}$) and outwardly increasing dust attenuation ($+0.056 \pm 0.015$\,mag\,kpc$^{-1}$) of SCGGe (Section~\ref{subsec:pixel}) are consistent with a centrally depleted, dust-poor inner region and a still star-forming, dust-enriched outer disk.
A caveat applies to SCGGe: if the western arc-like structure and south-southwest secondary component identified in the S\'ersic decomposition (Section~\ref{subsec:morphology}) represent a physically distinct merging satellite rather than tidal material associated with the primary galaxy, the measured radial profile may reflect a superposition of two stellar populations rather than an internal gradient within a single system. The inverted sSFR profile of SCGGe should therefore be interpreted as tentative, especially given its marginal Spearman significance ($p = 0.07$).

SCGGb, the most extended group member ($r_{1/2} = 1.93^{+0.74}_{-0.54}$\,kpc; Section~\ref{subsec:morphology}), shows a flat sSFR radial profile (Spearman $\rho = +0.68$, $p = 0.09$; Section~\ref{subsec:pixel}), with no evidence of centrally concentrated star formation. Its disk-like primary profile ($n = 0.77 \pm 0.10$) and elongated, asymmetric stellar distribution (two-component S\'ersic fit; Table~\ref{tab:sample}) suggest that its stellar light distribution is being reshaped by tidal forces from its close neighbours (SCGGc at 3.68\,pkpc; SCGGa at 3.19\,pkpc). The flat, extended sSFR distribution of SCGGb may reflect an interaction-driven redistribution of star-forming gas across a wider physical area, consistent with its morphologically disturbed appearance.

The group members thus display a diversity of resolved star-formation modes within a single $\sim\!16$\,pkpc system at $z \approx 5$: inside-out growth in three members, a tentative centrally suppressed mode in the most massive member, and tidally disturbed extended star formation in SCGGb. This suggests that the group environment is already differentiating the evolutionary trajectories of its members well before the system coalesces.

\subsection{Comparison to Known Analogues}
\label{subsec:analogues}

The most directly comparable system at a similar cosmic epoch is CGG-z5 at $z \sim 5.2$ \citep{Jin_et_al_2023}, six candidate members within a projected area of $\sim\!10 \times 20$\,kpc$^2$ in the EGS field. Its stellar mass range (most massive member $\log M_* \approx 9.8$, satellites $\log M_* \approx 8.4$--$9.2$) closely matches SCGG-z5. A critical distinction is that CGG-z5 is identified from photometric redshifts alone ($5.0 < z_{\rm phot} < 5.3$), whereas all six members of SCGG-z5 are spectroscopically confirmed via \ha\ emission. Within CGG-z5, high-resolution NIRCam imaging resolves the most massive member (CGG-z5.a) into three sub-components, and a second member (CGG-z5.d) into two, providing morphological evidence for ongoing mergers analogous to the tidal features and multi-component structure we identify in SCGG-z5. Given the comparable stellar mass range and projected scale of SCGG-z5, we expect a similar evolutionary outcome based on EAGLE simulations (Section~\ref{subsec:dynamics_discuss}).

CGG-z7 at $z \sim 7.04$ \citep{Wei_et_al_2026}, observed $\sim\!0.4$\,Gyr earlier, provides an earlier snapshot of a spectroscopically confirmed compact group, with four of six candidate members confirmed via [\ion{O}{3}] and H$\beta$ emission. The group spans $7.8 \times 5.7$\,kpc$^2$, with $\sigma_v = 93.7 \pm 31.7$\,\kms. As \citet{Wei_et_al_2026} note, the virial mass estimator is biased low because the system is not in dynamical equilibrium; CGG-z7 is interpreted as a pre-virialized structure near apocenter, and the apparent stellar-to-dynamical mass ratio ($M_*/M_{\rm vir} \approx 0.15$, a factor of $\sim\!3$ above typical virialized values) reflects non-virialized kinematics rather than a true equilibrium ratio. The much higher velocity dispersion of SCGG-z5 ($\sigma_v = 375^{+55}_{-195}$\,\kms) is consistent with a more dynamically evolved system in which multiple mutual passages have increased the internal kinetic energy. The central galaxy of CGG-z7 harbours a likely obscured AGN ([\ion{O}{3}]$\lambda$5007/H$\beta \approx 18$; \citealt{Wei_et_al_2026}), an outcome consistent with gas inflows driven by tidal interactions in compact group environments \citep{Ellison_et_al_2011}.
No AGN activity is identified within the compact core of SCGG-z5; the nearest broad-line AGN (ID\,11166; FWHM\,$= 2051 \pm 133$\,\kms; \citealt{Sun_et_al_2025}) lies $\approx\!30\arcsec$ ($\approx\!190$\,pkpc; $\Delta v \approx -383$\,\kms\ from the group mean) to the south, embedded within the same large-scale overdensity at $z \approx 4.97$ but well outside the group core, consistent with the tendency for AGN at $z \sim 4$--5 to preferentially trace overdense environments \citep{Lin_et_al_2024}.
In SCGG-z5, the ratio SFR$_{\mathrm{H}\alpha}$/SFR$_{\rm SED}$ varies widely across members (0.1--1.9). The most massive member, SCGGe, shows SFR$_{\mathrm{H}\alpha}$ $\approx 0.1\times$ SFR$_{\rm SED}$, consistent with its tentatively reduced central star formation and tentative inverted sSFR profile (Section~\ref{subsec:pixel}). SCGGa shows SFR$_{\mathrm{H}\alpha}$ $\approx 1.9\times$ SFR$_{\rm SED}$, possibly reflecting a recent star-formation episode not captured by the smooth delayed-$\tau$ SFH. The \textsc{Bagpipes} SED fitting further shows that SCGGe hosts the most evolved stellar population in the group: its best-fit SFH-onset age of $765^{+189}_{-175}$\,Myr is more than an order of magnitude older than the youngest actively star-forming members (SCGGd: $52^{+30}_{-21}$\,Myr; SCGGa: $65^{+35}_{-23}$\,Myr), consistent with an earlier onset of star formation in the most massive group member.

At a later epoch, CGG-z4 at $z = 4.3$ \citep{Brinch_et_al_2025} may represent a more advanced compact-group stage, with a combined IR star-formation rate of $\sim$2100\,$M_\odot$\,yr$^{-1}$ and gas depletion times of $\lesssim$100\,Myr for its two spectroscopically confirmed members, consistent with rapid consumption of the available gas reservoir. In this sense, SCGG-z5 may occupy an earlier evolutionary stage than CGG-z4, before the system reaches such extreme gas-rich, near-depletion conditions. Taken together, these systems suggest a plausible evolutionary progression from pre-virial first crossing (CGG-z7, $z \sim 7$), through dynamically active multi-passage group assembly (SCGG-z5, $z \approx 5$), to near-exhaustion of the star-forming gas reservoir (CGG-z4, $z \approx 4.3$).

A direct observational precedent for interaction-driven star-formation diversity in a compact group is provided by the A2744-Quintet at $z = 7.9$ \citep{Fudamoto_et_al_2025b}, a compact system of five merging galaxies within $\sim\!10$\,kpc in the core of the spectroscopically confirmed protocluster A2744-z7p9OD \citep{Morishita_et_al_2023}. Through JWST and ALMA [\ion{C}{2}]\,158\,$\mu$m observations, \citet{Fudamoto_et_al_2025b} directly trace tidal gas bridges connecting the merging members: galaxies stripped of their neutral gas reservoir show recent quenching, while those receiving the redistributed gas undergo starbursts, all on timescales of $\lesssim\!50$\,Myr. SCGG-z5 similarly exhibits a coexistence of suppressed star formation in its most massive member and actively star-forming companions spanning projected separations of $\sim\!5$--$13$\,pkpc, though the detailed configuration of SFH differentiation differs between the two systems. Future ALMA [\ion{C}{2}] observations of SCGG-z5 would directly test whether analogous merger-driven gas redistribution is responsible for the diversity of star-formation histories observed within the group.

Together, these systems illustrate that JWST is uncovering compact galaxy groups and overdensities already in place across a wide range of cosmic epochs \citep[e.g.,][]{Helton_et_al_2024b}. \citet{Morishita_et_al_2025a} identified two overdensities at $z \sim 5.7$ in the Abell~2744 field with an elevated fraction of evolved, weakly star-forming galaxies, pointing to an early onset of environmental effects on star formation. At $z \sim 6.9$, JWST/NIRSpec IFS of SPT0311-58 \citep{Arribas_et_al_2024} revealed 12 galaxies within $\sim\!17 \times 17$\,kpc$^2$ spanning diverse evolutionary stages and metallicities, providing a further example of compact multi-galaxy assembly in a dense proto-cluster core. Using the same SAPPHIRES-EDR data, \citet{Fudamoto_et_al_2025} discovered a proto-cluster candidate at $z = 8.47$ in the MACS0416 parallel field (nine spectroscopically confirmed members; $\delta \sim 6$--8 times the field average), further demonstrating that dense environments are already shaping galaxy evolution at the earliest cosmic epochs.

A complementary view of a more advanced assembly stage is offered by the ``Crimson Behemoth'' CID-931 at $z = 4.91$ \citep{Tanaka_et_al_2024}, a dusty AGN host surrounded by $\geq\!8$ massive star-forming clumps ($\log M_* \sim 9$--$10$ each) within $\sim\!10$\,kpc. \citet{Tanaka_et_al_2024} interpret this morphology as either a complex merger of multiple massive galaxies or violent disk instability. Pixel-by-pixel SED fitting of CID-931 recovers significant spatial variation between clumps, with circumnuclear regions younger and more bursty than the central body, providing a resolved comparison to our \textsc{Bagpipes} analysis. If SCGG-z5 represents the pre-coalescence stage at which individual members are still spectroscopically distinguishable, CID-931 may illustrate
the subsequent interpenetrating phase once members begin to merge: the interaction-triggered AGN at its centre is consistent with gas inflows expected from compact, gas-rich pairs at small projected separations
\citep{Ellison_et_al_2011}.

\section{Summary}
\label{sec:summary}

We have discovered and characterised SCGG-z5 (the Shirui Group), a compact galaxy proto-group at $z \approx 4.97$ in the MACS0416 field, using data from the SAPPHIRES Early Data Release \citep{Sun_et_al_2025}.

\begin{enumerate}

\item Six galaxies are spectroscopically confirmed as group members via \ha\ emission, spanning $4.958 \leq z_{\rm spec} \leq 4.979$ within a projected diameter of $\sim$16.1\,pkpc. The overdensity corresponds to a $7.1\sigma$ spike in the field \ha-emitter distribution (Poisson $p = 6.8 \times 10^{-13}$).

\item Individual stellar masses span
$8.42 \leq \log(M_*/\msun) \leq 9.84$, with a total group stellar mass of $\log(M_*/\msun) = 10.07\pm 0.04$ and a combined \ha-based SFR of $\sim$25\,$\msun$\,yr$^{-1}$ (Theios+2019 calibration). Three of the six members lie above or on the \citet{Di_Cesare_et_al_2026}
H$\alpha$-based SFMS at $z\sim4$--5, by up to $0.5$\,dex.

\item All six members show disturbed morphologies in F200W, with multi-component substructure identified in S\'ersic fitting (an arc-like feature in SCGGe consistent with tidal disturbance; compact two-component structure in SCGGd and SCGGa; an elongated, asymmetric stellar distribution in SCGGb) and bipolar residual patterns in three of the six members. The residual signal is outer-dominated across the full group, suggestive of group-wide tidal perturbations. Pixel-by-pixel SED fitting is consistent with inside-out stellar mass growth in three members (SCGGf, SCGGd, SCGGa; $\Delta\log\,\mathrm{sSFR} = 0.43$--$0.44$\,dex); SCGGc shows a statistically significant but much weaker gradient ($\Delta\log\,\mathrm{sSFR} = 0.07$\,dex) with an integrated SFR within the intrinsic scatter of the SFMS ($\Delta{\rm MS} = -0.28$\,dex, $|\Delta{\rm MS}| < \sigma_{\rm int}$); the weak gradient amplitude is inconsistent with the pronounced inside-out growth seen in SCGGf, SCGGd, and SCGGa. The most massive member (SCGGe, $\log M_*/\msun = 9.84^{+0.04}_{-0.05}$) shows a tentative inverted sSFR profile suggestive of reduced central star formation \citep{Looser_et_al_2025}.

\item The velocity dispersion over all six members is $\sigma_v = 375^{+55}_{-195}$\,\kms, with a dynamical crossing time $t_{\rm cross} \approx 14.1$\,Myr ($\sim$1.2\,per\,cent of the cosmic age at this epoch). The projected mass estimator yields $M_{\rm PM} = 2.0 \times 10^{12}$\,$\msun$ ($\log M_{\rm PM}/\msun \approx 12.30^{+0.30}_{-0.25}$), with a stellar fraction of $\approx$0.57\,per\,cent, consistent with a dark-matter-dominated group halo.

\item By analogy with EAGLE simulations of similar structures \citep{Jin_et_al_2023}, SCGG-z5 is likely to coalesce into a single galaxy by $z \sim 3$--$4$ and assemble $\log(M_*/\msun) > 11$ by $z \sim 1$, consistent with a short-lived, pre-coalescence phase of massive 
galaxy formation.

\end{enumerate}

SCGG-z5 is a rare example of a spectroscopically confirmed compact galaxy proto-group at $z > 4$, offering direct observational constraints on early group assembly during the first gigayear of cosmic history. The diversity of star-formation histories and morphological disturbance observed across its six members, embedded within a $7.1\sigma$ galaxy overdensity, is consistent with environmental effects already differentiating the evolutionary trajectories of member galaxies well within the first gigayear of cosmic time. Future JWST/NIRSpec integral-field spectroscopy would simultaneously map the spatially resolved H$\alpha$ and [O\,\textsc{iii}]\,$\lambda\lambda$4959,5007 kinematics, gas-phase metallicity, and ionised-gas diagnostics across all six members, directly probing the dynamical state of the group and the role of tidal interactions in driving the inside-out star formation identified here. Complementary ALMA observations of the [C\,\textsc{ii}]\,158\,$\mu$m line and dust continuum would further constrain the cold gas distribution, total gas reservoir, and dust-obscured star formation across the group, providing a more complete view of its baryonic content at this early epoch.

\section*{Data Availability}
The specific JWST data presented in this article were obtained from the Mikulski Archive for Space Telescopes (MAST) at the Space Telescope Science Institute. The specific observations analysed can be accessed via \dataset[doi:10.17909/0w8h-jt25]{https://doi.org/10.17909/0w8h-jt25}. The reduced data are also publicly available via the SAPPHIRES survey website at \url{https://jwst-sapphires.github.io}.

\begin{acknowledgments}
We thank the anonymous referee for a thorough and constructive review that improved the manuscript.
This work was supported by JSPS KAKENHI grant Nos. 23H01219 and 26H02070, and JSPS Core-to-Core Program (grant Nos.: JPJSCCA2021003 and JPJSCCA20260002). H.K. acknowledges support from JSPS KAKENHI grant Nos. 23KJ2148 and 25K17444. T.K. and Y.K. acknowledge financial support from JSPS KAKENHI Grant Numbers 24H00002 (Specially Promoted Research by T. Kodama et al.), 22K21349 (International Leading Research by S. Miyazaki et al.).

This work is based on observations made with the NASA/ESA/CSA James Webb Space Telescope. The data were obtained from the Mikulski Archive for Space Telescopes at the Space Telescope Science Institute, which is operated by the Association of Universities for Research in Astronomy, Inc., under NASA contract NAS 5-03127 for JWST. These observations are associated with program \#6434.
Support for program \#6434 was provided by NASA through a grant from the Space Telescope Science Institute, which is operated by the Association of Universities for Research in Astronomy, Inc., under NASA contract NAS 5-03127.
\end{acknowledgments}

\bibliography{sample701}{}

@ARTICLE{MendesdeOliveira_Hickson_1994,
       author = {{Mendes de Oliveira}, Claudia and {Hickson}, Paul},
        title = "{Morphology of Galaxies in Compact Groups}",
      journal = {\apj},
         year = 1994,
        month = jun,
       volume = {427},
        pages = {684},
          doi = {10.1086/174175},
       adsurl = {https://ui.adsabs.harvard.edu/abs/1994ApJ...427..684M}
}

@ARTICLE{Cluver_et_al_2013,
       author = {{Cluver}, M.~E. and {Appleton}, P.~N. and {Ogle}, P. and {Jarrett}, T.~H. and {Rasmussen}, J. and {Lisenfeld}, U. and {Guillard}, P. and {Verdes-Montenegro}, L. and {Antonucci}, R. and {Bitsakis}, T. and {Charmandaris}, V. and {Boulanger}, F. and {Egami}, E. and {Xu}, C.~K. and {Yun}, M.~S.},
        title = "{Enhanced Warm H$_{2}$ Emission in the Compact Group Mid-infrared ``Green Valley''}",
      journal = {\apj},
         year = 2013,
        month = mar,
       volume = {765},
       number = {2},
          eid = {93},
        pages = {93},
          doi = {10.1088/0004-637X/765/2/93},
archivePrefix = {arXiv},
       eprint = {1301.4549},
 primaryClass = {astro-ph.CO},
       adsurl = {https://ui.adsabs.harvard.edu/abs/2013ApJ...765...93C}
}

@ARTICLE{Jin_et_al_2023,
       author = {{Jin}, Shuowen and {Sillassen}, Nikolaj B. and {Magdis}, Georgios E. and {Vijayan}, Aswin P. and {Brammer}, Gabriel B. and {Kokorev}, Vasily and {Weaver}, John R. and {Gobat}, Raphael and {Gim{\'e}nez-Arteaga}, Clara and {Valentino}, Francesco and {Brinch}, Malte and {G{\'o}mez-Guijarro}, Carlos and {Shuntov}, Marko and {Toft}, Sune and {Greve}, Thomas R. and {Blanquez Sese}, David},
        title = "{Massive galaxy formation caught in action at z {\ensuremath{\sim}} 5 with JWST}",
      journal = {\aap},
         year = 2023,
        month = feb,
       volume = {670},
          eid = {L11},
        pages = {L11},
          doi = {10.1051/0004-6361/202245724},
archivePrefix = {arXiv},
       eprint = {2212.09372},
 primaryClass = {astro-ph.GA},
       adsurl = {https://ui.adsabs.harvard.edu/abs/2023A&A...670L..11J}
}

@ARTICLE{Brinch_et_al_2025,
       author = {{Brinch}, Malte and {Jin}, Shuowen and {Gobat}, Raphael and {Sillassen}, Nikolaj B. and {Algera}, Hiddo and {Gillman}, Steven and {Greve}, Thomas R. and {Gomez-Guijarro}, Carlos and {Gullberg}, Bitten and {Hodge}, Jacqueline and {Lee}, Minju and {Liu}, Daizhong and {Magdis}, Georgios and {Valentino}, Francesco},
        title = "{Revealing the hidden cosmic feast: A z = 4.3 galaxy group hosting two optically dark, efficiently star-forming galaxies}",
      journal = {\aap},
         year = 2025,
        month = feb,
       volume = {694},
          eid = {A218},
        pages = {A218},
          doi = {10.1051/0004-6361/202451448},
archivePrefix = {arXiv},
       eprint = {2501.05288},
 primaryClass = {astro-ph.GA},
       adsurl = {https://ui.adsabs.harvard.edu/abs/2025A&A...694A.218B}
}

@ARTICLE{Wei_et_al_2026,
       author = {{Wei}, Xiaoyang and {Cai}, Zheng and {Yu}, Fujiang and {Li}, Mingyu and {Wu}, Yunjing},
        title = "{Pre-Virialized Assembly at Cosmic Dawn: The Dynamics and Extreme Ionization of Compact Group CGG-z7 at $z\sim7.04$}",
      journal = {arXiv e-prints},
         year = 2026,
        month = mar,
          eid = {arXiv:2603.08066},
        pages = {arXiv:2603.08066},
          doi = {10.48550/arXiv.2603.08066},
archivePrefix = {arXiv},
       eprint = {2603.08066},
 primaryClass = {astro-ph.GA},
       adsurl = {https://ui.adsabs.harvard.edu/abs/2026arXiv260308066W}
}

@ARTICLE{Tanaka_et_al_2024,
       author = {{Tanaka}, Takumi S. and {Silverman}, John D. and {Nakazato}, Yurina and {Onoue}, Masafusa and {Shimasaku}, Kazuhiro and {Fudamoto}, Yoshinobu and {Fujimoto}, Seiji and {Ding}, Xuheng and {Faisst}, Andreas L. and {Valentino}, Francesco and {Jin}, Shuowen and {Hayward}, Christopher C. and {Kokorev}, Vasily and {Ceverino}, Daniel and {Kalita}, Boris S. and {Casey}, Caitlin M. and {Liu}, Zhaoxuan and {Kaminsky}, Aidan and {Fei}, Qinyue and {Andika}, Irham T. and {Lambrides}, Erini and {Akins}, Hollis B. and {Kartaltepe}, Jeyhan S. and {Koekemoer}, Anton M. and {McCracken}, Henry Joy and {Rhodes}, Jason and {Robertson}, Brant E. and {Franco}, Maximilien and {Liu}, Daizhong and {Chartab}, Nima and {Gillman}, Steven and {Gozaliasl}, Ghassem and {Hirschmann}, Michaela and {Huertas-Company}, Marc and {Massey}, Richard and {Roy}, Namrata and {Sattari}, Zahra and {Shuntov}, Marko and {Sterling}, Joseph and {Toft}, Sune and {Trakhtenbrot}, Benny and {Yoshida}, Naoki and {Zavala}, Jorge A.},
        title = "{Crimson Behemoth: A massive clumpy structure hosting a dusty AGN at z=4.91}",
      journal = {\pasj},
         year = 2024,
        month = dec,
       volume = {76},
       number = {6},
        pages = {1323-1335},
          doi = {10.1093/pasj/psae091},
archivePrefix = {arXiv},
       eprint = {2410.00104},
 primaryClass = {astro-ph.GA},
       adsurl = {https://ui.adsabs.harvard.edu/abs/2024PASJ...76.1323T}
}

@ARTICLE{Sun_et_al_2025,
       author = {{Sun}, Fengwu and {Fudamoto}, Yoshinobu and {Lin}, Xiaojing and {Helton}, Jakob M. and {Hsiao}, Tiger Yu-Yang and {Egami}, Eiichi and {Akhtarkavan}, Arshia and {Bunker}, Andrew J. and {Cai}, Zheng and {DeCoursey}, Christa and {Eisenstein}, Daniel J. and {Fan}, Xiaohui and {Harikane}, Yuichi and {Ji}, Zhiyuan and {Jin}, Xiangyu and {Liu}, Weizhe and {Liu}, Yichen and {Ma}, Zheng and {Maiolino}, Roberto and {Ouchi}, Masami and {Tee}, Wei Leong and {Wang}, Feige and {Willmer}, Christopher N.~A. and {Wu}, Yunjing and {Xu}, Yi and {Yang}, Jinyi and {Zhang}, Junyu and {Zhu}, Yongda},
        title = "{Slitless Areal Pure-Parallel HIgh-Redshift Emission Survey (SAPPHIRES): Early Data Release of Deep JWST/NIRCam Images and Spectra in MACS J0416 Parallel Field}",
      journal = {arXiv e-prints},
         year = 2025,
        month = mar,
          eid = {arXiv:2503.15587},
        pages = {arXiv:2503.15587},
          doi = {10.48550/arXiv.2503.15587},
archivePrefix = {arXiv},
       eprint = {2503.15587},
 primaryClass = {astro-ph.GA},
       adsurl = {https://ui.adsabs.harvard.edu/abs/2025arXiv250315587S}
}

@ARTICLE{Hopkins_et_al_2009,
       author = {{Hopkins}, Philip F. and {Cox}, Thomas J. and {Younger}, Joshua D. and {Hernquist}, Lars},
        title = "{How do Disks Survive Mergers?}",
      journal = {\apj},
         year = 2009,
        month = feb,
       volume = {691},
       number = {2},
        pages = {1168-1201},
          doi = {10.1088/0004-637X/691/2/1168},
archivePrefix = {arXiv},
       eprint = {0806.1739},
 primaryClass = {astro-ph},
       adsurl = {https://ui.adsabs.harvard.edu/abs/2009ApJ...691.1168H}
}

@ARTICLE{Oser_et_al_2010,
       author = {{Oser}, Ludwig and {Ostriker}, Jeremiah P. and {Naab}, Thorsten and {Johansson}, Peter H. and {Burkert}, Andreas},
        title = "{The Two Phases of Galaxy Formation}",
      journal = {\apj},
         year = 2010,
        month = dec,
       volume = {725},
       number = {2},
        pages = {2312-2323},
          doi = {10.1088/0004-637X/725/2/2312},
archivePrefix = {arXiv},
       eprint = {1010.1381},
 primaryClass = {astro-ph.CO},
       adsurl = {https://ui.adsabs.harvard.edu/abs/2010ApJ...725.2312O}
}

@ARTICLE{Hirschmann_et_al_2012,
       author = {{Hirschmann}, Michaela and {Naab}, Thorsten and {Somerville}, Rachel S. and {Burkert}, Andreas and {Oser}, Ludwig},
        title = "{Galaxy formation in semi-analytic models and cosmological hydrodynamic zoom simulations}",
      journal = {\mnras},
         year = 2012,
        month = feb,
       volume = {419},
       number = {4},
        pages = {3200-3222},
          doi = {10.1111/j.1365-2966.2011.19961.x},
archivePrefix = {arXiv},
       eprint = {1104.1626},
 primaryClass = {astro-ph.CO},
       adsurl = {https://ui.adsabs.harvard.edu/abs/2012MNRAS.419.3200H}
}

@ARTICLE{Overzier_2016,
       author = {{Overzier}, Roderik A.},
        title = "{The realm of the galaxy protoclusters. A review}",
      journal = {\aapr},
         year = 2016,
        month = nov,
       volume = {24},
       number = {1},
          eid = {14},
        pages = {14},
          doi = {10.1007/s00159-016-0100-3},
archivePrefix = {arXiv},
       eprint = {1610.05201},
 primaryClass = {astro-ph.GA},
       adsurl = {https://ui.adsabs.harvard.edu/abs/2016A&ARv..24...14O}
}

@ARTICLE{Chiang_et_al_2017,
       author = {{Chiang}, Yi-Kuan and {Overzier}, Roderik A. and {Gebhardt}, Karl and {Henriques}, Bruno},
        title = "{Galaxy Protoclusters as Drivers of Cosmic Star Formation History in the First 2 Gyr}",
      journal = {\apjl},
         year = 2017,
        month = aug,
       volume = {844},
       number = {2},
          eid = {L23},
        pages = {L23},
          doi = {10.3847/2041-8213/aa7e7b},
archivePrefix = {arXiv},
       eprint = {1705.01634},
 primaryClass = {astro-ph.GA},
       adsurl = {https://ui.adsabs.harvard.edu/abs/2017ApJ...844L..23C}
}

@ARTICLE{Marrone_et_al_2018,
       author = {{Marrone}, D.~P. and {Spilker}, J.~S. and {Hayward}, C.~C. and {Vieira}, J.~D. and {Aravena}, M. and {Ashby}, M.~L.~N. and {Bayliss}, M.~B. and {B{\'e}thermin}, M. and {Brodwin}, M. and {Bothwell}, M.~S. and {Carlstrom}, J.~E. and {Chapman}, S.~C. and {Chen}, Chian-Chou and {Crawford}, T.~M. and {Cunningham}, D.~J.~M. and {De Breuck}, C. and {Fassnacht}, C.~D. and {Gonzalez}, A.~H. and {Greve}, T.~R. and {Hezaveh}, Y.~D. and {Lacaille}, K. and {Litke}, K.~C. and {Lower}, S. and {Ma}, J. and {Malkan}, M. and {Miller}, T.~B. and {Morningstar}, W.~R. and {Murphy}, E.~J. and {Narayanan}, D. and {Phadke}, K.~A. and {Rotermund}, K.~M. and {Sreevani}, J. and {Stalder}, B. and {Stark}, A.~A. and {Strandet}, M.~L. and {Tang}, M. and {Wei{\ss}}, A.},
        title = "{Galaxy growth in a massive halo in the first billion years of cosmic history}",
      journal = {\nat},
         year = 2018,
        month = jan,
       volume = {553},
       number = {7686},
        pages = {51-54},
          doi = {10.1038/nature24629},
archivePrefix = {arXiv},
       eprint = {1712.03020},
 primaryClass = {astro-ph.GA},
       adsurl = {https://ui.adsabs.harvard.edu/abs/2018Natur.553...51M}
}

@ARTICLE{Dekel_Burkert_2014,
       author = {{Dekel}, A. and {Burkert}, A.},
        title = "{Wet disc contraction to galactic blue nuggets and quenching to red nuggets}",
      journal = {\mnras},
         year = 2014,
        month = feb,
       volume = {438},
       number = {2},
        pages = {1870-1879},
          doi = {10.1093/mnras/stt2331},
archivePrefix = {arXiv},
       eprint = {1310.1074},
 primaryClass = {astro-ph.CO},
       adsurl = {https://ui.adsabs.harvard.edu/abs/2014MNRAS.438.1870D}
}

@ARTICLE{Tacchella_et_al_2016,
       author = {{Tacchella}, Sandro and {Dekel}, Avishai and {Carollo}, C. Marcella and {Ceverino}, Daniel and {DeGraf}, Colin and {Lapiner}, Sharon and {Mandelker}, Nir and {Primack}, Joel R.},
        title = "{Evolution of density profiles in high-z galaxies: compaction and quenching inside-out}",
      journal = {\mnras},
         year = 2016,
        month = may,
       volume = {458},
       number = {1},
        pages = {242-263},
          doi = {10.1093/mnras/stw303},
archivePrefix = {arXiv},
       eprint = {1509.00017},
 primaryClass = {astro-ph.GA},
       adsurl = {https://ui.adsabs.harvard.edu/abs/2016MNRAS.458..242T}
}

@ARTICLE{Hopkins_et_al_2011,
       author = {{Hopkins}, Philip F. and {Quataert}, Eliot and {Murray}, Norman},
        title = "{Self-regulated star formation in galaxies via momentum input from massive stars}",
      journal = {\mnras},
         year = 2011,
        month = oct,
       volume = {417},
       number = {2},
        pages = {950-973},
          doi = {10.1111/j.1365-2966.2011.19306.x},
archivePrefix = {arXiv},
       eprint = {1101.4940},
 primaryClass = {astro-ph.CO},
       adsurl = {https://ui.adsabs.harvard.edu/abs/2011MNRAS.417..950H}
}

@ARTICLE{Ellison_et_al_2011,
       author = {{Ellison}, Sara L. and {Patton}, David R. and {Mendel}, J. Trevor and {Scudder}, Jillian M.},
        title = "{Galaxy pairs in the Sloan Digital Sky Survey - IV. Interactions trigger active galactic nuclei}",
      journal = {\mnras},
         year = 2011,
        month = dec,
       volume = {418},
       number = {3},
        pages = {2043-2053},
          doi = {10.1111/j.1365-2966.2011.19624.x},
archivePrefix = {arXiv},
       eprint = {1108.2711},
 primaryClass = {astro-ph.CO},
       adsurl = {https://ui.adsabs.harvard.edu/abs/2011MNRAS.418.2043E}
}

@ARTICLE{Ando_et_al_2022,
       author = {{Ando}, Makoto and {Shimasaku}, Kazuhiro and {Momose}, Rieko and {Ito}, Kei and {Sawicki}, Marcin and {Shimakawa}, Rhythm},
        title = "{A systematic search for galaxy protocluster cores at the transition epoch of their star formation activity}",
      journal = {\mnras},
         year = 2022,
        month = jul,
       volume = {513},
       number = {3},
        pages = {3252-3272},
          doi = {10.1093/mnras/stac1049},
archivePrefix = {arXiv},
       eprint = {2201.05185},
 primaryClass = {astro-ph.GA},
       adsurl = {https://ui.adsabs.harvard.edu/abs/2022MNRAS.513.3252A}
}

@ARTICLE{Girelli_et_al_2019,
       author = {{Girelli}, Giacomo and {Bolzonella}, Micol and {Cimatti}, Andrea},
        title = "{Massive and old quiescent galaxies at high redshift}",
      journal = {\aap},
         year = 2019,
        month = dec,
       volume = {632},
          eid = {A80},
        pages = {A80},
          doi = {10.1051/0004-6361/201834547},
archivePrefix = {arXiv},
       eprint = {1910.07544},
 primaryClass = {astro-ph.GA},
       adsurl = {https://ui.adsabs.harvard.edu/abs/2019A&A...632A..80G}
}

@ARTICLE{Carnall_et_al_2023,
       author = {{Carnall}, Adam C. and {McLure}, Ross J. and {Dunlop}, James S. and {McLeod}, Derek J. and {Wild}, Vivienne and {Cullen}, Fergus and {Magee}, Dan and {Begley}, Ryan and {Cimatti}, Andrea and {Donnan}, Callum T. and {Hamadouche}, Massissilia L. and {Jewell}, Sophie M. and {Walker}, Sam},
        title = "{A massive quiescent galaxy at redshift 4.658}",
      journal = {\nat},
         year = 2023,
        month = jul,
       volume = {619},
       number = {7971},
        pages = {716-719},
          doi = {10.1038/s41586-023-06158-6},
archivePrefix = {arXiv},
       eprint = {2301.11413},
 primaryClass = {astro-ph.GA},
       adsurl = {https://ui.adsabs.harvard.edu/abs/2023Natur.619..716C}
}

@ARTICLE{Jin_et_al_2024,
       author = {{Jin}, Shuowen and {Sillassen}, Nikolaj B. and {Magdis}, Georgios E. and {Brinch}, Malte and {Shuntov}, Marko and {Brammer}, Gabriel and {Gobat}, Raphael and {Valentino}, Francesco and {Carnall}, Adam C. and {Lee}, Minju and {Vijayan}, Aswin P. and {Gillman}, Steven and {Kokorev}, Vasily and {Le Bail}, Aur{\'e}lien and {Greve}, Thomas R. and {Gullberg}, Bitten and {Gould}, Katriona M.~L. and {Toft}, Sune},
        title = "{Cosmic Vine: A z = 3.44 large-scale structure hosting massive quiescent galaxies}",
      journal = {\aap},
         year = 2024,
        month = mar,
       volume = {683},
          eid = {L4},
        pages = {L4},
          doi = {10.1051/0004-6361/202348540},
archivePrefix = {arXiv},
       eprint = {2311.04867},
 primaryClass = {astro-ph.GA},
       adsurl = {https://ui.adsabs.harvard.edu/abs/2024A&A...683L...4J}
}

@ARTICLE{Diaz-Santos_et_al_2018,
       author = {{D{\'\i}az-Santos}, T. and {Assef}, R.~J. and {Blain}, A.~W. and {Aravena}, M. and {Stern}, D. and {Tsai}, C.-W. and {Eisenhardt}, P. and {Wu}, J. and {Jun}, H.~D. and {Dibert}, K. and {Inami}, H. and {Lansbury}, G. and {Leclercq}, F.},
        title = "{The multiple merger assembly of a hyperluminous obscured quasar at redshift 4.6}",
      journal = {Science},
         year = 2018,
        month = nov,
       volume = {362},
       number = {6418},
        pages = {1034-1036},
          doi = {10.1126/science.aap7605},
archivePrefix = {arXiv},
       eprint = {1811.05992},
 primaryClass = {astro-ph.GA},
       adsurl = {https://ui.adsabs.harvard.edu/abs/2018Sci...362.1034D}
}

@ARTICLE{Sillassen_et_al_2022,
       author = {{Sillassen}, Nikolaj B. and {Jin}, Shuowen and {Magdis}, Georgios E. and {Daddi}, Emanuele and {Weaver}, John R. and {Gobat}, Raphael and {Kokorev}, Vasily and {Valentino}, Francesco and {Finoguenov}, Alexis and {Shuntov}, Marko and {G{\'o}mez-Guijarro}, Carlos and {Coogan}, Rosemary and {Greve}, Thomas R. and {Toft}, Sune and {Blanquez Sese}, David},
        title = "{A galaxy group candidate at z {\ensuremath{\approx}} 3.7 in the COSMOS field}",
      journal = {\aap},
         year = 2022,
        month = sep,
       volume = {665},
          eid = {L7},
        pages = {L7},
          doi = {10.1051/0004-6361/202244661},
archivePrefix = {arXiv},
       eprint = {2209.05895},
 primaryClass = {astro-ph.GA},
       adsurl = {https://ui.adsabs.harvard.edu/abs/2022A&A...665L...7S}
}

@ARTICLE{Laishram_et_al_2026a,
       author = {{Laishram}, Ronaldo and {Koyama}, Yusei and {Naufal}, Abdurrahman and {Kodama}, Tadayuki and {Shimakawa}, Rhythm and {Daikuhara}, Kazuki and {Dannerbauer}, Helmut and {P{\'e}rez-Mart{\'\i}nez}, Jose Manuel and {P{\'e}rez-Gonz{\'a}lez}, Pablo G.},
        title = "{Spider-webb: Spatially Resolved Evidence of Inside-out Quenching in the Spiderweb Protocluster at z {\ensuremath{\sim}} 2}",
      journal = {\apj},
         year = 2026,
        month = feb,
       volume = {998},
       number = {1},
          eid = {158},
        pages = {158},
          doi = {10.3847/1538-4357/ae3003},
archivePrefix = {arXiv},
       eprint = {2512.18805},
 primaryClass = {astro-ph.GA},
       adsurl = {https://ui.adsabs.harvard.edu/abs/2026ApJ...998..158L}
}

@ARTICLE{Sillassen_et_al_2024,
       author = {{Sillassen}, Nikolaj B. and {Jin}, Shuowen and {Magdis}, Georgios E. and {Daddi}, Emanuele and {Wang}, Tao and {Lu}, Shiying and {Sun}, Hanwen and {Arumugam}, Vinod and {Liu}, Daizhong and {Brinch}, Malte and {D'Eugenio}, Chiara and {Gobat}, Raphael and {G{\'o}mez-Guijarro}, Carlos and {Rich}, Michael and {Schinnerer}, Eva and {Strazzullo}, Veronica and {Tan}, Qinghua and {Valentino}, Francesco and {Wang}, Yijun and {Xiao}, Mengyuan and {Zhou}, Luwenjia and {Bl{\'a}nquez-Ses{\'e}}, David and {Cai}, Zheng and {Chen}, Yanmei and {Ciesla}, Laure and {Dai}, Yu and {Delvecchio}, Ivan and {Elbaz}, David and {Finoguenov}, Alexis and {Gao}, Fangyou and {Gu}, Qiusheng and {Hale}, Catherine and {Hao}, Qiaoyang and {Huang}, Jiasheng and {Jarvis}, Matt and {Kalita}, Boris and {Ke}, Xu and {Le Bail}, Aurelien and {Magnelli}, Benjamin and {Shi}, Yong and {Vaccari}, Mattia and {Whittam}, Imogen and {Yang}, Tiancheng and {Zhang}, Zhiyu},
        title = "{NOEMA formIng Cluster survEy (NICE): Characterizing eight massive galaxy groups at 1.5 < z < 4 in the COSMOS field}",
      journal = {\aap},
         year = 2024,
        month = oct,
       volume = {690},
          eid = {A55},
        pages = {A55},
          doi = {10.1051/0004-6361/202450760},
archivePrefix = {arXiv},
       eprint = {2407.02973},
 primaryClass = {astro-ph.GA},
       adsurl = {https://ui.adsabs.harvard.edu/abs/2024A&A...690A..55S}
}

@ARTICLE{Fudamoto_et_al_2025,
       author = {{Fudamoto}, Yoshinobu and {Nakazato}, Yurina and {Ceverino}, Daniel and {Colina}, Luis and {Hashimoto}, Takuya and {Inoue}, Akio K. and {Tamura}, Yoichi and {Yoshida}, Naoki and {Zhu}, Yongda and {Sugahara}, Yuma and {Arribas}, Santiago and {'Arvarez-M'arquez}, Javier and {Bakx}, Tom and {Blanco Prieto}, Carmen and {Costantin}, Luca and {Crespo G'omez}, Alejandro and {Hagimoto}, Masato and {Hashigaya}, Takeshi and {Matsuo}, Hiroshi and {Marques-Chaves}, Rui and {Mawatari}, Ken and {Mitsuhashi}, Ikki and {Osone}, Wataru and {Pereira-Santaella}, Miguel and {Umehata}, Hideki and {Witten}, Callum and {Ren}, Yi W.},
        title = "{Early massive galaxy formation in the core of a galaxy protocluster 650 million years after the Big Bang}",
      journal = {arXiv e-prints},
         year = 2025,
        month = oct,
          eid = {arXiv:2510.11770},
        pages = {arXiv:2510.11770},
          doi = {10.48550/arXiv.2510.11770},
archivePrefix = {arXiv},
       eprint = {2510.11770},
 primaryClass = {astro-ph.GA},
       adsurl = {https://ui.adsabs.harvard.edu/abs/2025arXiv251011770F}
}

@ARTICLE{Carnall_et_al_2018,
       author = {{Carnall}, A.~C. and {McLure}, R.~J. and {Dunlop}, J.~S. and {Dav{\'e}}, R.},
        title = "{Inferring the star formation histories of massive quiescent galaxies with BAGPIPES: evidence for multiple quenching mechanisms}",
      journal = {\mnras},
         year = 2018,
        month = nov,
       volume = {480},
       number = {4},
        pages = {4379-4401},
          doi = {10.1093/mnras/sty2169},
archivePrefix = {arXiv},
       eprint = {1712.04452},
 primaryClass = {astro-ph.GA},
       adsurl = {https://ui.adsabs.harvard.edu/abs/2018MNRAS.480.4379C}
}

@ARTICLE{Lange_2023,
       author = {{Lange}, Johannes U.},
        title = "{NAUTILUS: boosting Bayesian importance nested sampling with deep learning}",
      journal = {\mnras},
         year = 2023,
        month = oct,
       volume = {525},
       number = {2},
        pages = {3181-3194},
          doi = {10.1093/mnras/stad2441},
archivePrefix = {arXiv},
       eprint = {2306.16923},
 primaryClass = {astro-ph.IM},
       adsurl = {https://ui.adsabs.harvard.edu/abs/2023MNRAS.525.3181L}
}

@ARTICLE{Kroupa_1993,
       author = {{Kroupa}, Pavel and {Tout}, Christopher A. and {Gilmore}, Gerard},
        title = "{The Distribution of Low-Mass Stars in the Galactic Disc}",
      journal = {\mnras},
         year = 1993,
        month = jun,
       volume = {262},
        pages = {545-587},
          doi = {10.1093/mnras/262.3.545},
       adsurl = {https://ui.adsabs.harvard.edu/abs/1993MNRAS.262..545K}
}

@ARTICLE{Stanway_Eldridge_2018,
       author = {{Stanway}, E.~R. and {Eldridge}, J.~J.},
        title = "{Re-evaluating old stellar populations}",
      journal = {\mnras},
         year = 2018,
        month = sep,
       volume = {479},
       number = {1},
        pages = {75-93},
          doi = {10.1093/mnras/sty1353},
archivePrefix = {arXiv},
       eprint = {1805.08784},
 primaryClass = {astro-ph.GA},
       adsurl = {https://ui.adsabs.harvard.edu/abs/2018MNRAS.479...75S}
}

@ARTICLE{Hsiao_et_al_2023,
       author = {{Hsiao}, Tiger Yu-Yang and {Coe}, Dan and {Abdurro'uf} and {Whitler}, Lily and {Jung}, Intae and {Khullar}, Gourav and {Meena}, Ashish Kumar and {Dayal}, Pratika and {Barrow}, Kirk S.~S. and {Santos-Olmsted}, Lillian and {Casselman}, Adam and {Vanzella}, Eros and {Nonino}, Mario and {Jim{\'e}nez-Teja}, Yolanda and {Oguri}, Masamune and {Stark}, Daniel P. and {Furtak}, Lukas J. and {Zitrin}, Adi and {Adamo}, Angela and {Brammer}, Gabriel and {Bradley}, Larry and {Diego}, Jose M. and {Zackrisson}, Erik and {Finkelstein}, Steven L. and {Windhorst}, Rogier A. and {Bhatawdekar}, Rachana and {Hutchison}, Taylor A. and {Broadhurst}, Tom and {Dimauro}, Paola and {Andrade-Santos}, Felipe and {Eldridge}, Jan J. and {Acebron}, Ana and {Avila}, Roberto J. and {Bayliss}, Matthew B. and {Ben{\'\i}tez}, Alex and {Binggeli}, Christian and {Bolan}, Patricia and {Brada{\v{c}}}, Maru{\v{s}}a and {Carnall}, Adam C. and {Conselice}, Christopher J. and {Donahue}, Megan and {Frye}, Brenda and {Fujimoto}, Seiji and {Henry}, Alaina and {James}, Bethan L. and {Kassin}, Susan A. and {Kewley}, Lisa and {Larson}, Rebecca L. and {Lauer}, Tod and {Law}, David and {Mahler}, Guillaume and {Mainali}, Ramesh and {McCandliss}, Stephan and {Nicholls}, David and {Pirzkal}, Norbert and {Postman}, Marc and {Rigby}, Jane R. and {Ryan}, Russell and {Senchyna}, Peter and {Sharon}, Keren and {Shimizu}, Ikko and {Strait}, Victoria and {Tang}, Mengtao and {Trenti}, Michele and {Vikaeus}, Anton and {Welch}, Brian},
        title = "{JWST Reveals a Possible z {\ensuremath{\sim}} 11 Galaxy Merger in Triply Lensed MACS0647─JD}",
      journal = {\apjl},
         year = 2023,
        month = jun,
       volume = {949},
       number = {2},
          eid = {L34},
        pages = {L34},
          doi = {10.3847/2041-8213/acc94b},
archivePrefix = {arXiv},
       eprint = {2210.14123},
 primaryClass = {astro-ph.GA},
       adsurl = {https://ui.adsabs.harvard.edu/abs/2023ApJ...949L..34H}
}

@ARTICLE{Calzetti_et_al_2000,
       author = {{Calzetti}, Daniela and {Armus}, Lee and {Bohlin}, Ralph C. and {Kinney}, Anne L. and {Koornneef}, Jan and {Storchi-Bergmann}, Thaisa},
        title = "{The Dust Content and Opacity of Actively Star-forming Galaxies}",
      journal = {\apj},
         year = 2000,
        month = apr,
       volume = {533},
       number = {2},
        pages = {682-695},
          doi = {10.1086/308692},
archivePrefix = {arXiv},
       eprint = {astro-ph/9911459},
 primaryClass = {astro-ph},
       adsurl = {https://ui.adsabs.harvard.edu/abs/2000ApJ...533..682C}
}

@ARTICLE{Pasha_Miller_2023,
       author = {{Pasha}, Imad and {Miller}, Tim B.},
        title = "{pysersic: A Python package for determining galaxy structural properties via Bayesian inference, accelerated with jax}",
      journal = {The Journal of Open Source Software},
         year = 2023,
        month = sep,
       volume = {8},
       number = {89},
          eid = {5703},
        pages = {5703},
          doi = {10.21105/joss.05703},
archivePrefix = {arXiv},
       eprint = {2306.05454},
 primaryClass = {astro-ph.GA},
       adsurl = {https://ui.adsabs.harvard.edu/abs/2023JOSS....8.5703P}
}

@ARTICLE{Abdurrouf_et_al_2021,
       author = {{Abdurro'uf} and {Lin}, Yen-Ting and {Wu}, Po-Feng and {Akiyama}, Masayuki},
        title = "{Introducing piXedfit: A Spectral Energy Distribution Fitting Code Designed for Resolved Sources}",
      journal = {\apjs},
         year = 2021,
        month = may,
       volume = {254},
       number = {1},
          eid = {15},
        pages = {15},
          doi = {10.3847/1538-4365/abebe2},
archivePrefix = {arXiv},
       eprint = {2101.09717},
 primaryClass = {astro-ph.GA},
       adsurl = {https://ui.adsabs.harvard.edu/abs/2021ApJS..254...15A}
}

@ARTICLE{Behroozi_et_al_2019,
       author = {{Behroozi}, Peter and {Wechsler}, Risa H. and {Hearin}, Andrew P. and {Conroy}, Charlie},
        title = "{UNIVERSEMACHINE: The correlation between galaxy growth and dark matter halo assembly from z = 0-10}",
      journal = {\mnras},
         year = 2019,
        month = sep,
       volume = {488},
       number = {3},
        pages = {3143-3194},
          doi = {10.1093/mnras/stz1182},
archivePrefix = {arXiv},
       eprint = {1806.07893},
 primaryClass = {astro-ph.GA},
       adsurl = {https://ui.adsabs.harvard.edu/abs/2019MNRAS.488.3143B}
}

@ARTICLE{Heisler_et_al_1985,
       author = {{Heisler}, J. and {Tremaine}, S. and {Bahcall}, J.~N.},
        title = "{Estimating the masses of galaxy groups: alternatives to the virial theorem.}",
      journal = {\apj},
         year = 1985,
        month = nov,
       volume = {298},
        pages = {8-17},
          doi = {10.1086/163584},
       adsurl = {https://ui.adsabs.harvard.edu/abs/1985ApJ...298....8H}
}

@ARTICLE{Tacchella_et_al_2018,
       author = {{Tacchella}, S. and {Carollo}, C.~M. and {F{\"o}rster Schreiber}, N.~M. and {Renzini}, A. and {Dekel}, A. and {Genzel}, R. and {Lang}, P. and {Lilly}, S.~J. and {Mancini}, C. and {Onodera}, M. and {Tacconi}, L.~J. and {Wuyts}, S. and {Zamorani}, G.},
        title = "{Dust Attenuation, Bulge Formation, and Inside-out Quenching of Star Formation in Star-forming Main Sequence Galaxies at z {\ensuremath{\sim}} 2}",
      journal = {\apj},
         year = 2018,
        month = may,
       volume = {859},
       number = {1},
          eid = {56},
        pages = {56},
          doi = {10.3847/1538-4357/aabf8b},
archivePrefix = {arXiv},
       eprint = {1704.00733},
 primaryClass = {astro-ph.GA},
       adsurl = {https://ui.adsabs.harvard.edu/abs/2018ApJ...859...56T}
}

@ARTICLE{Barrera-Ballesteros_et_al_2015,
       author = {{Barrera-Ballesteros}, J.~K. and {S{\'a}nchez}, S.~F. and {Garc{\'\i}a-Lorenzo}, B. and {Falc{\'o}n-Barroso}, J. and {Mast}, D. and {Garc{\'\i}a-Benito}, R. and {Husemann}, B. and {van de Ven}, G. and {Iglesias-P{\'a}ramo}, J. and {Rosales-Ortega}, F.~F. and {P{\'e}rez-Torres}, M.~A. and {M{\'a}rquez}, I. and {Kehrig}, C. and {Marino}, R.~A. and {Vilchez}, J.~M. and {Galbany}, L. and {L{\'o}pez-S{\'a}nchez}, {\'A}. R. and {Walcher}, C.~J. and {CALIFA Collaboration}},
        title = "{Central star formation and metallicity in CALIFA interacting galaxies}",
      journal = {\aap},
         year = 2015,
        month = jul,
       volume = {579},
          eid = {A45},
        pages = {A45},
          doi = {10.1051/0004-6361/201425397},
archivePrefix = {arXiv},
       eprint = {1505.03153},
 primaryClass = {astro-ph.GA},
       adsurl = {https://ui.adsabs.harvard.edu/abs/2015A&A...579A..45B}
}

@ARTICLE{Spilker_et_al_2019,
       author = {{Spilker}, Justin S. and {Bezanson}, Rachel and {Weiner}, Benjamin J. and {Whitaker}, Katherine E. and {Williams}, Christina C.},
        title = "{Evidence for Inside-out Galaxy Growth and Quenching of a z {\ensuremath{\sim}} 2 Compact Galaxy From High-resolution Molecular Gas Imaging}",
      journal = {\apj},
         year = 2019,
        month = sep,
       volume = {883},
       number = {1},
          eid = {81},
        pages = {81},
          doi = {10.3847/1538-4357/ab3804},
archivePrefix = {arXiv},
       eprint = {1908.02294},
 primaryClass = {astro-ph.GA},
       adsurl = {https://ui.adsabs.harvard.edu/abs/2019ApJ...883...81S}
}

@ARTICLE{Calzetti_2001,
       author = {{Calzetti}, Daniela},
        title = "{The Dust Opacity of Star-forming Galaxies}",
      journal = {\pasp},
         year = 2001,
        month = dec,
       volume = {113},
       number = {790},
        pages = {1449-1485},
          doi = {10.1086/324269},
archivePrefix = {arXiv},
       eprint = {astro-ph/0109035},
 primaryClass = {astro-ph},
       adsurl = {https://ui.adsabs.harvard.edu/abs/2001PASP..113.1449C}
}

@ARTICLE{Valentino_et_al_2020,
       author = {{Valentino}, Francesco and {Tanaka}, Masayuki and {Davidzon}, Iary and {Toft}, Sune and {G{\'o}mez-Guijarro}, Carlos and {Stockmann}, Mikkel and {Onodera}, Masato and {Brammer}, Gabriel and {Ceverino}, Daniel and {Faisst}, Andreas L. and {Gallazzi}, Anna and {Hayward}, Christopher C. and {Ilbert}, Olivier and {Kubo}, Mariko and {Magdis}, Georgios E. and {Selsing}, Jonatan and {Shimakawa}, Rhythm and {Sparre}, Martin and {Steinhardt}, Charles and {Yabe}, Kiyoto and {Zabl}, Johannes},
        title = "{Quiescent Galaxies 1.5 Billion Years after the Big Bang and Their Progenitors}",
      journal = {\apj},
         year = 2020,
        month = feb,
       volume = {889},
       number = {2},
          eid = {93},
        pages = {93},
          doi = {10.3847/1538-4357/ab64dc},
archivePrefix = {arXiv},
       eprint = {1909.10540},
 primaryClass = {astro-ph.GA},
       adsurl = {https://ui.adsabs.harvard.edu/abs/2020ApJ...889...93V}
}

@ARTICLE{Morishita_et_al_2023,
       author = {{Morishita}, Takahiro and {Roberts-Borsani}, Guido and {Treu}, Tommaso and {Brammer}, Gabriel and {Mason}, Charlotte A. and {Trenti}, Michele and {Vulcani}, Benedetta and {Wang}, Xin and {Acebron}, Ana and {Bah{\'e}}, Yannick and {Bergamini}, Pietro and {Boyett}, Kristan and {Bradac}, Marusa and {Calabr{\`o}}, Antonello and {Castellano}, Marco and {Chen}, Wenlei and {De Lucia}, Gabriella and {Filippenko}, Alexei V. and {Fontana}, Adriano and {Glazebrook}, Karl and {Grillo}, Claudio and {Henry}, Alaina and {Jones}, Tucker and {Kelly}, Patrick L. and {Koekemoer}, Anton M. and {Leethochawalit}, Nicha and {Lu}, Ting-Yi and {Marchesini}, Danilo and {Mascia}, Sara and {Mercurio}, Amata and {Merlin}, Emiliano and {Metha}, Benjamin and {Nanayakkara}, Themiya and {Nonino}, Mario and {Paris}, Diego and {Pentericci}, Laura and {Rosati}, Piero and {Santini}, Paola and {Strait}, Victoria and {Vanzella}, Eros and {Windhorst}, Rogier A. and {Xie}, Lizhi},
        title = "{Early Results from GLASS-JWST. XIV. A Spectroscopically Confirmed Protocluster 650 Million Years after the Big Bang}",
      journal = {\apjl},
         year = 2023,
        month = apr,
       volume = {947},
       number = {2},
          eid = {L24},
        pages = {L24},
          doi = {10.3847/2041-8213/acb99e},
archivePrefix = {arXiv},
       eprint = {2211.09097},
 primaryClass = {astro-ph.GA},
       adsurl = {https://ui.adsabs.harvard.edu/abs/2023ApJ...947L..24M}
}

@ARTICLE{Morishita_et_al_2025a,
       author = {{Morishita}, Takahiro and {Liu}, Zhaoran and {Stiavelli}, Massimo and {Treu}, Tommaso and {Trenti}, Michele and {Chartab}, Nima and {Roberts-Borsani}, Guido and {Vulcani}, Benedetta and {Bergamini}, Pietro and {Castellano}, Marco and {Grillo}, Claudio},
        title = "{Accelerated Emergence of Evolved Galaxies in Early Overdensities at z {\ensuremath{\sim}} 5.7}",
      journal = {\apj},
         year = 2025,
        month = apr,
       volume = {982},
       number = {2},
          eid = {153},
        pages = {153},
          doi = {10.3847/1538-4357/adb30f},
archivePrefix = {arXiv},
       eprint = {2408.10980},
 primaryClass = {astro-ph.GA},
       adsurl = {https://ui.adsabs.harvard.edu/abs/2025ApJ...982..153M}
}

@ARTICLE{Di_Cesare_et_al_2026,
       author = {{Di Cesare}, Claudia and {Matthee}, Jorryt and {Naidu}, Rohan P. and {Torralba}, Alberto and {Kotiwale}, Gauri and {Kramarenko}, Ivan G. and {Blaizot}, Jeremy and {Rosdahl}, Joakim and {Leja}, Joel and {Iani}, Edoardo and {Adamo}, Angela and {Covelo-Paz}, Alba and {Furtak}, Lukas J. and {Heintz}, Kasper E. and {Mascia}, Sara and {Navarrete}, Benjam{\'\i}n and {Oesch}, Pascal A. and {Romano}, Michael and {Shivaei}, Irene and {Tacchella}, Sandro},
        title = "{The slope and scatter of the star-forming main sequence at z {\ensuremath{\sim}} 5: Reconciling observations with simulations}",
      journal = {\aap},
         year = 2026,
        month = mar,
       volume = {707},
          eid = {A129},
        pages = {A129},
          doi = {10.1051/0004-6361/202557790},
archivePrefix = {arXiv},
       eprint = {2510.19044},
 primaryClass = {astro-ph.GA},
       adsurl = {https://ui.adsabs.harvard.edu/abs/2026A&A...707A.129D}
}

@ARTICLE{Theios_et_al_2019,
       author = {{Theios}, Rachel L. and {Steidel}, Charles C. and {Strom}, Allison L. and {Rudie}, Gwen C. and {Trainor}, Ryan F. and {Reddy}, Naveen A.},
        title = "{Dust Attenuation, Star Formation, and Metallicity in z {\ensuremath{\sim}} 2-3 Galaxies from KBSS-MOSFIRE}",
      journal = {\apj},
         year = 2019,
        month = jan,
       volume = {871},
       number = {1},
          eid = {128},
        pages = {128},
          doi = {10.3847/1538-4357/aaf386},
archivePrefix = {arXiv},
       eprint = {1805.00016},
 primaryClass = {astro-ph.GA},
       adsurl = {https://ui.adsabs.harvard.edu/abs/2019ApJ...871..128T}
}

@ARTICLE{Looser_et_al_2025,
       author = {{Looser}, Tobias J. and {D'Eugenio}, Francesco and {Maiolino}, Roberto and {Tacchella}, Sandro and {Curti}, Mirko and {Arribas}, Santiago and {Baker}, William M. and {Baum}, Stefi and {Bonaventura}, Nina and {Boyett}, Kristan and {Bunker}, Andrew J. and {Carniani}, Stefano and {Charlot}, Stephane and {Chevallard}, Jacopo and {Curtis-Lake}, Emma and {Lola Danhaive}, A. and {Eisenstein}, Daniel J. and {de Graaff}, Anna and {Hainline}, Kevin and {Ji}, Zhiyuan and {Johnson}, Benjamin D. and {Kumari}, Nimisha and {Nelson}, Erica and {Parlanti}, Eleonora and {Rix}, Hans-Walter and {Robertson}, Brant and {Del Pino}, Bruno Rodr{\'\i}guez and {Sandles}, Lester and {Scholtz}, Jan and {Smit}, Renske and {Stark}, Daniel P. and {{\"U}bler}, Hannah and {Williams}, Christina C. and {Willott}, Chris and {Witstok}, Joris},
        title = "{JADES: Differing assembly histories of galaxies: Observational evidence for bursty star formation histories and (mini-)quenching in the first billion years of the Universe}",
      journal = {\aap},
         year = 2025,
        month = may,
       volume = {697},
          eid = {A88},
        pages = {A88},
          doi = {10.1051/0004-6361/202347102},
archivePrefix = {arXiv},
       eprint = {2306.02470},
 primaryClass = {astro-ph.GA},
       adsurl = {https://ui.adsabs.harvard.edu/abs/2025A&A...697A..88L}
}

@ARTICLE{Fudamoto_et_al_2025b,
       author = {{Fudamoto}, Yoshinobu and {Nakazato}, Yurina and {Ceverino}, Daniel and {Colina}, Luis and {Hashimoto}, Takuya and {Inoue}, Akio K. and {Tamura}, Yoichi and {Yoshida}, Naoki and {Zhu}, Yongda and {Sugahara}, Yuma and {Arribas}, Santiago and {'Arvarez-M'arquez}, Javier and {Bakx}, Tom and {Blanco Prieto}, Carmen and {Costantin}, Luca and {Crespo G'omez}, Alejandro and {Hagimoto}, Masato and {Hashigaya}, Takeshi and {Matsuo}, Hiroshi and {Marques-Chaves}, Rui and {Mawatari}, Ken and {Mitsuhashi}, Ikki and {Osone}, Wataru and {Pereira-Santaella}, Miguel and {Umehata}, Hideki and {Witten}, Callum and {Ren}, Yi W.},
        title = "{Early massive galaxy formation in the core of a galaxy protocluster 650 million years after the Big Bang}",
      journal = {arXiv e-prints},
         year = 2025,
        month = oct,
          eid = {arXiv:2510.11770},
        pages = {arXiv:2510.11770},
          doi = {10.48550/arXiv.2510.11770},
archivePrefix = {arXiv},
       eprint = {2510.11770},
 primaryClass = {astro-ph.GA},
       adsurl = {https://ui.adsabs.harvard.edu/abs/2025arXiv251011770F}
}

@ARTICLE{Lin_et_al_2024,
       author = {{Lin}, Xiaojing and {Wang}, Feige and {Fan}, Xiaohui and {Cai}, Zheng and {Champagne}, Jaclyn B. and {Sun}, Fengwu and {Volonteri}, Marta and {Yang}, Jinyi and {Hennawi}, Joseph F. and {Ba{\~n}ados}, Eduardo and {Barth}, Aaron and {Eilers}, Anna-Christina and {Farina}, Emanuele Paolo and {Liu}, Weizhe and {Jin}, Xiangyu and {Jun}, Hyunsung D. and {Lupi}, Alessandro and {Kakiichi}, Koki and {Mazzucchelli}, Chiara and {Onoue}, Masafusa and {Pan}, Zhiwei and {Pizzati}, Elia and {Rojas-Ruiz}, Sof{\'\i}a and {Schindler}, Jan-Torge and {Trakhtenbrot}, Benny and {Shen}, Yue and {Trebitsch}, Maxime and {Zhuang}, Ming-Yang and {Endsley}, Ryan and {Meyer}, Romain A. and {Li}, Zihao and {Li}, Mingyu and {Pudoka}, Maria and {Tee}, Wei Leong and {Wu}, Yunjing and {Zhang}, Haowen},
        title = "{A SPectroscopic Survey of Biased Halos In the Reionization Era (ASPIRE): Broad-line AGN at z = 4{\ensuremath{-}}5 Revealed by JWST/NIRCam WFSS}",
      journal = {\apj},
         year = 2024,
        month = oct,
       volume = {974},
       number = {1},
          eid = {147},
        pages = {147},
          doi = {10.3847/1538-4357/ad6565},
archivePrefix = {arXiv},
       eprint = {2407.17570},
 primaryClass = {astro-ph.GA},
       adsurl = {https://ui.adsabs.harvard.edu/abs/2024ApJ...974..147L}
}

@ARTICLE{Helton_et_al_2024a,
       author = {{Helton}, Jakob M. and {Sun}, Fengwu and {Woodrum}, Charity and {Hainline}, Kevin N. and {Willmer}, Christopher N.~A. and {Rieke}, George H. and {Rieke}, Marcia J. and {Tacchella}, Sandro and {Robertson}, Brant and {Johnson}, Benjamin D. and {Alberts}, Stacey and {Eisenstein}, Daniel J. and {Hausen}, Ryan and {Bonaventura}, Nina R. and {Bunker}, Andrew and {Charlot}, Stephane and {Curti}, Mirko and {Curtis-Lake}, Emma and {Looser}, Tobias J. and {Maiolino}, Roberto and {Willott}, Chris and {Witstok}, Joris and {Boyett}, Kristan and {Chen}, Zuyi and {Egami}, Eiichi and {Endsley}, Ryan and {Hviding}, Raphael E. and {Jaffe}, Daniel T. and {Ji}, Zhiyuan and {Lyu}, Jianwei and {Sandles}, Lester},
        title = "{The JWST Advanced Deep Extragalactic Survey: Discovery of an Extreme Galaxy Overdensity at z = 5.4 with JWST/NIRCam in GOODS-S}",
      journal = {\apj},
         year = 2024,
        month = feb,
       volume = {962},
       number = {2},
          eid = {124},
        pages = {124},
          doi = {10.3847/1538-4357/ad0da7},
archivePrefix = {arXiv},
       eprint = {2302.10217},
 primaryClass = {astro-ph.GA},
       adsurl = {https://ui.adsabs.harvard.edu/abs/2024ApJ...962..124H}
}

@ARTICLE{Helton_et_al_2024b,
       author = {{Helton}, Jakob M. and {Sun}, Fengwu and {Woodrum}, Charity and {Hainline}, Kevin N. and {Willmer}, Christopher N.~A. and {Rieke}, Marcia J. and {Rieke}, George H. and {Alberts}, Stacey and {Eisenstein}, Daniel J. and {Tacchella}, Sandro and {Robertson}, Brant and {Johnson}, Benjamin D. and {Baker}, William M. and {Bhatawdekar}, Rachana and {Bunker}, Andrew J. and {Chen}, Zuyi and {Egami}, Eiichi and {Ji}, Zhiyuan and {Maiolino}, Roberto and {Willott}, Chris and {Witstok}, Joris},
        title = "{Identification of High-redshift Galaxy Overdensities in GOODS-N and GOODS-S}",
      journal = {\apj},
         year = 2024,
        month = oct,
       volume = {974},
       number = {1},
          eid = {41},
        pages = {41},
          doi = {10.3847/1538-4357/ad6867},
archivePrefix = {arXiv},
       eprint = {2311.04270},
 primaryClass = {astro-ph.GA},
       adsurl = {https://ui.adsabs.harvard.edu/abs/2024ApJ...974...41H}
}

@ARTICLE{Arribas_et_al_2024,
       author = {{Arribas}, Santiago and {Perna}, Michele and {Rodr{\'\i}guez Del Pino}, Bruno and {Lamperti}, Isabella and {D'Eugenio}, Francesco and {P{\'e}rez-Gonz{\'a}lez}, Pablo G. and {Jones}, Gareth C. and {Crespo G{\'o}mez}, Alejandro and {Curti}, Mirko and {Lim}, Seunghwan and {{\'A}lvarez-M{\'a}rquez}, Javier and {Bunker}, Andrew J. and {Carniani}, Stefano and {Charlot}, St{\'e}phane and {Jakobsen}, Peter and {Maiolino}, Roberto and {{\"U}bler}, Hannah and {Willott}, Chris J. and {B{\"o}ker}, Torsten and {Chevallard}, Jacopo and {Circosta}, Chiara and {Cresci}, Giovanni and {Kumari}, Nimisha and {Parlanti}, Eleonora and {Scholtz}, Jan and {Venturi}, Giacomo and {Witstok}, Joris},
        title = "{GA-NIFS: The core of an extremely massive protocluster at the epoch of reionisation probed with JWST/NIRSpec}",
      journal = {\aap},
         year = 2024,
        month = aug,
       volume = {688},
          eid = {A146},
        pages = {A146},
          doi = {10.1051/0004-6361/202348824},
archivePrefix = {arXiv},
       eprint = {2312.00899},
 primaryClass = {astro-ph.GA},
       adsurl = {https://ui.adsabs.harvard.edu/abs/2024A&A...688A.146A}
}

@ARTICLE{Morishita_et_al_2025b,
       author = {{Morishita}, Takahiro and {Stiavelli}, Massimo and {Vanzella}, Eros and {Bergamini}, Pietro and {Boyett}, Kristan and {Chiaberge}, Marco and {Grillo}, Claudio and {Leethochawalit}, Nicha and {Messa}, Matteo and {Roberts-Borsani}, Guido and {Rosati}, Piero and {Shajib}, Anowar J.},
        title = "{Metallicity Scatter Originating from Subkiloparsec Starbursting Clumps in the Core of a Protocluster at z = 7.88}",
      journal = {\apj},
         year = 2025,
        month = may,
       volume = {985},
       number = {1},
          eid = {83},
        pages = {83},
          doi = {10.3847/1538-4357/adc4c3},
archivePrefix = {arXiv},
       eprint = {2501.11879},
 primaryClass = {astro-ph.GA},
       adsurl = {https://ui.adsabs.harvard.edu/abs/2025ApJ...985...83M}
}

@ARTICLE{Sun_et_al_2024,
       author = {{Sun}, Fengwu and {Helton}, Jakob M. and {Egami}, Eiichi and {Hainline}, Kevin N. and {Rieke}, George H. and {Willmer}, Christopher N.~A. and {Eisenstein}, Daniel J. and {Johnson}, Benjamin D. and {Rieke}, Marcia J. and {Robertson}, Brant and {Tacchella}, Sandro and {Alberts}, Stacey and {Baker}, William M. and {Bhatawdekar}, Rachana and {Boyett}, Kristan and {Bunker}, Andrew J. and {Charlot}, Stephane and {Chen}, Zuyi and {Chevallard}, Jacopo and {Curtis-Lake}, Emma and {Danhaive}, A. Lola and {DeCoursey}, Christa and {Ji}, Zhiyuan and {Lyu}, Jianwei and {Maiolino}, Roberto and {Rujopakarn}, Wiphu and {Sandles}, Lester and {Shivaei}, Irene and {{\"U}bler}, Hannah and {Willott}, Chris and {Witstok}, Joris},
        title = "{JADES: Resolving the Stellar Component and Filamentary Overdense Environment of Hubble Space Telescope (HST)-dark Submillimeter Galaxy HDF850.1 at z = 5.18}",
      journal = {\apj},
         year = 2024,
        month = jan,
       volume = {961},
       number = {1},
          eid = {69},
        pages = {69},
          doi = {10.3847/1538-4357/ad07e3},
archivePrefix = {arXiv},
       eprint = {2309.04529},
 primaryClass = {astro-ph.GA},
       adsurl = {https://ui.adsabs.harvard.edu/abs/2024ApJ...961...69S}
}

@ARTICLE{Haryana_et_al_2025,
       author = {{Haryana}, Novan Saputra and {Akiyama}, Masayuki and {Abdurro'uf} and {Wulandari}, Hesti Retno Tri and {Alfonzo}, Juan Pablo and {Lee}, Kianhong and {Matsumoto}, Naoki and {Sutanto}, Ryo Albert and {Effendi}, Muhammad Nur Ihsan and {Fitriana}, Itsna Khoirul and {Huda}, Ibnu Nurul and {Jaelani}, Anton Timur and {Kusuma}, Sultan Hadi and {Puspitarini}, Lucky and {Triani}, Dian P.},
        title = "{Stellar Mass Assembly History of Massive Quiescent Galaxies since z {\ensuremath{\sim}} 4: Insights from Spatially Resolved Spectral Energy Distribution Fitting with JWST Data}",
      journal = {\apj},
         year = 2025,
        month = dec,
       volume = {994},
       number = {2},
          eid = {215},
        pages = {215},
          doi = {10.3847/1538-4357/ae03ad},
archivePrefix = {arXiv},
       eprint = {2508.19170},
 primaryClass = {astro-ph.GA},
       adsurl = {https://ui.adsabs.harvard.edu/abs/2025ApJ...994..215H}
}

@ARTICLE{Valentino_et_al_2023,
       author = {{Valentino}, Francesco and {Brammer}, Gabriel and {Gould}, Katriona M.~L. and {Kokorev}, Vasily and {Fujimoto}, Seiji and {Jespersen}, Christian Kragh and {Vijayan}, Aswin P. and {Weaver}, John R. and {Ito}, Kei and {Tanaka}, Masayuki and et al.},
        title = "{An Atlas of Color-selected Quiescent Galaxies at z > 3 in Public JWST Fields}",
      journal = {\apj},
         year = 2023,
        month = apr,
       volume = {947},
       number = {1},
          eid = {20},
        pages = {20},
          doi = {10.3847/1538-4357/acbefa},
archivePrefix = {arXiv},
       eprint = {2302.10936},
 primaryClass = {astro-ph.GA},
       adsurl = {https://ui.adsabs.harvard.edu/abs/2023ApJ...947...20V}
}

@ARTICLE{Baker_et_al_2025,
       author = {{Baker}, William M. and {Valentino}, Francesco and {Lagos}, Claudia del P. and {Ito}, Kei and {Jespersen}, Christian Kragh and {Gottumukkala}, Rashmi and {Hjorth}, Jens and {Langeroodi}, Danial and {Sedgewick}, Aidan},
        title = "{Exploring over 700 massive quiescent galaxies at z = 2─7: Demographics and stellar mass functions}",
      journal = {\aap},
         year = 2025,
        month = oct,
       volume = {702},
          eid = {A270},
        pages = {A270},
          doi = {10.1051/0004-6361/202555829},
archivePrefix = {arXiv},
       eprint = {2506.04119},
 primaryClass = {astro-ph.GA},
       adsurl = {https://ui.adsabs.harvard.edu/abs/2025A&A...702A.270B}
}

@ARTICLE{Laishram_et_al_2026b,
       author = {{Laishram}, Ronaldo and {Koyama}, Yusei and {Kusakabe}, Haruka and {Kikuta}, Satoshi and {Shimizu}, Shunta and {Kodama}, Tadayuki},
        title = "{Discovery of a z ≃ 4.9 Ly{\ensuremath{\alpha}} Emitter Protocluster: Wavelength-dependent Environmental Effects on Galaxy Structure}",
      journal = {\apjl},
         year = 2026,
        month = may,
       volume = {1002},
       number = {1},
          eid = {L25},
        pages = {L25},
          doi = {10.3847/2041-8213/ae5824},
archivePrefix = {arXiv},
       eprint = {2603.03570},
 primaryClass = {astro-ph.GA},
       adsurl = {https://ui.adsabs.harvard.edu/abs/2026ApJ..1002L..25L}
}

@ARTICLE{Ginolfi_et_al_2020,
       author = {{Ginolfi}, M. and {Jones}, G.~C. and {B{\'e}thermin}, M. and {Faisst}, A. and {Lemaux}, B.~C. and {Schaerer}, D. and {Fudamoto}, Y. and {Oesch}, P. and {Dessauges-Zavadsky}, M. and {Fujimoto}, S. and {Carniani}, S. and {Le F{\`e}vre}, O. and {Cassata}, P. and {Silverman}, J.~D. and {Capak}, P. and {Yan}, Lin and {Bardelli}, S. and {Cucciati}, O. and {Gal}, R. and {Gruppioni}, C. and {Hathi}, N.~P. and {Lubin}, L. and {Maiolino}, R. and {Morselli}, L. and {Pelliccia}, D. and {Talia}, M. and {Vergani}, D. and {Zamorani}, G.},
        title = "{The ALPINE-ALMA [CII] survey. Circumgalactic medium pollution and gas mixing by tidal stripping in a merging system at z {\ensuremath{\sim}} 4.57}",
      journal = {\aap},
         year = 2020,
        month = nov,
       volume = {643},
          eid = {A7},
        pages = {A7},
          doi = {10.1051/0004-6361/202038284},
archivePrefix = {arXiv},
       eprint = {2004.13737},
 primaryClass = {astro-ph.GA},
       adsurl = {https://ui.adsabs.harvard.edu/abs/2020A&A...643A...7G}
}

@ARTICLE{Pozzetti_et_al_2010,
       author = {{Pozzetti}, L. and {Bolzonella}, M. and {Zucca}, E. and {Zamorani}, G. and {Lilly}, S. and {Renzini}, A. and {Moresco}, M. and {Mignoli}, M. and {Cassata}, P. and {Tasca}, L. and {Lamareille}, F. and {Maier}, C. and {Meneux}, B. and {Halliday}, C. and {Oesch}, P. and {Vergani}, D. and {Caputi}, K. and {Kova{\v{c}}}, K. and {Cimatti}, A. and {Cucciati}, O. and {Iovino}, A. and {Peng}, Y. and {Carollo}, M. and {Contini}, T. and {Kneib}, J.-P. and {Le F{\'e}vre}, O. and {Mainieri}, V. and {Scodeggio}, M. and {Bardelli}, S. and {Bongiorno}, A. and {Coppa}, G. and {de la Torre}, S. and {de Ravel}, L. and {Franzetti}, P. and {Garilli}, B. and {Kampczyk}, P. and {Knobel}, C. and {Le Borgne}, J.-F. and {Le Brun}, V. and {Pell{\`o}}, R. and {Perez Montero}, E. and {Ricciardelli}, E. and {Silverman}, J.~D. and {Tanaka}, M. and {Tresse}, L. and {Abbas}, U. and {Bottini}, D. and {Cappi}, A. and {Guzzo}, L. and {Koekemoer}, A.~M. and {Leauthaud}, A. and {Maccagni}, D. and {Marinoni}, C. and {McCracken}, H.~J. and {Memeo}, P. and {Porciani}, C. and {Scaramella}, R. and {Scarlata}, C. and {Scoville}, N.},
        title = "{zCOSMOS - 10k-bright spectroscopic sample. The bimodality in the galaxy stellar mass function: exploring its evolution with redshift}",
      journal = {\aap},
         year = 2010,
        month = nov,
       volume = {523},
          eid = {A13},
        pages = {A13},
          doi = {10.1051/0004-6361/200913020},
archivePrefix = {arXiv},
       eprint = {0907.5416},
 primaryClass = {astro-ph.CO},
       adsurl = {https://ui.adsabs.harvard.edu/abs/2010A&A...523A..13P}
}

@ARTICLE{Paquereau_et_al_2025,
       author = {{Paquereau}, L. and {Laigle}, C. and {McCracken}, H.~J. and {Shuntov}, M. and {Ilbert}, O. and {Akins}, H.~B. and {Allen}, N. and {Arango-Togo}, R. and {Berman}, E.~M. and {B{\'e}thermin}, M. and {Casey}, C.~M. and {McCleary}, J. and {Dubois}, Y. and {Drakos}, N.~E. and {Faisst}, A.~L. and {Franco}, M. and {Harish}, S. and {Jespersen}, C.~K. and {Kartaltepe}, J.~S. and {Koekemoer}, A.~M. and {Kokorev}, V. and {Lambrides}, E. and {Larson}, R. and {Liu}, D. and {Le Borgne}, D. and {Lewis}, J.~S.~W. and {McKinney}, J. and {Mercier}, W. and {Rhodes}, J.~D. and {Robertson}, B.~E. and {Toft}, S. and {Trebitsch}, M. and {Tresse}, L. and {Weaver}, J.~R.},
        title = "{Tracing the galaxy-halo connection with galaxy clustering in COSMOS-Web from z = 0.1 to z {\ensuremath{\sim}} 12}",
      journal = {\aap},
         year = 2025,
        month = oct,
       volume = {702},
          eid = {A163},
        pages = {A163},
          doi = {10.1051/0004-6361/202553828},
archivePrefix = {arXiv},
       eprint = {2501.11674},
 primaryClass = {astro-ph.GA},
       adsurl = {https://ui.adsabs.harvard.edu/abs/2025A&A...702A.163P}
}

@ARTICLE{Evrard_et_al_2008,
       author = {{Evrard}, A.~E. and {Bialek}, J. and {Busha}, M. and {White}, M. and {Habib}, S. and {Heitmann}, K. and {Warren}, M. and {Rasia}, E. and {Tormen}, G. and {Moscardini}, L. and {Power}, C. and {Jenkins}, A.~R. and {Gao}, L. and {Frenk}, C.~S. and {Springel}, V. and {White}, S.~D.~M. and {Diemand}, J.},
        title = "{Virial Scaling of Massive Dark Matter Halos: Why Clusters Prefer a High Normalization Cosmology}",
      journal = {\apj},
         year = 2008,
        month = jan,
       volume = {672},
       number = {1},
        pages = {122-137},
          doi = {10.1086/521616},
archivePrefix = {arXiv},
       eprint = {astro-ph/0702241},
 primaryClass = {astro-ph},
       adsurl = {https://ui.adsabs.harvard.edu/abs/2008ApJ...672..122E}
}

@ARTICLE{Shuntov_et_al_2025,
       author = {{Shuntov}, M. and {Ilbert}, O. and {Toft}, S. and {Arango-Toro}, R.~C. and {Akins}, H.~B. and {Casey}, C.~M. and {Franco}, M. and {Harish}, S. and {Kartaltepe}, J.~S. and {Koekemoer}, A.~M. and {McCracken}, H.~J. and {Paquereau}, L. and {Laigle}, C. and {Bethermin}, M. and {Dubois}, Y. and {Drakos}, N.~E. and {Faisst}, A. and {Gozaliasl}, G. and {Gillman}, S. and {Hayward}, C.~C. and {Hirschmann}, M. and {Huertas-Company}, M. and {Jespersen}, C.~K. and {Jin}, S. and {Kokorev}, V. and {Lambrides}, E. and {Le Borgne}, D. and {Liu}, D. and {Magdis}, G. and {Massey}, R. and {McPartland}, C.~J.~R. and {Mercier}, W. and {McCleary}, J.~E. and {McKinney}, J. and {Oesch}, P.~A. and {Renzini}, A. and {Rhodes}, J.~D. and {Rich}, R.~M. and {Robertson}, B.~E. and {Sanders}, D. and {Trebitsch}, M. and {Tresse}, L. and {Valentino}, F. and {Vijayan}, A.~P. and {Weaver}, J.~R. and {Weibel}, A. and {Wilkins}, S.~M. and {Yang}, L.},
        title = "{COSMOS-Web: Stellar mass assembly in relation to dark matter halos across 0.2 < z < 12 of cosmic history}",
      journal = {\aap},
         year = 2025,
        month = mar,
       volume = {695},
          eid = {A20},
        pages = {A20},
          doi = {10.1051/0004-6361/202452570},
archivePrefix = {arXiv},
       eprint = {2410.08290},
 primaryClass = {astro-ph.GA},
       adsurl = {https://ui.adsabs.harvard.edu/abs/2025A&A...695A..20S}
}
\bibliographystyle{aasjournalv7}

\end{document}